\documentclass[useAMS,usenatbib,usegraphicx]{mn2e}
\newcommand{\cmsq}{cm$^{-2}$}
\newcommand{\cmt}{cm$^{-3}$}
\newcommand{\pc}{{\rm pc}}

\newcommand{\msun}{M$_\odot$}
\newcommand{\msunpc}{M$_\odot$ pc$^{-2}$}
\newcommand{\zsun}{Z$_\odot$}
\newcommand{\kms}{km s$^{-1}$}
\newcommand{\Kkms}{K km s$^{-1}$}
\newcommand{\Xunits}{cm$^{-2}$ K$^{-1}$ km$^{-1}$ s}
\newcommand{\Wunits}{K km s$^{-1}$}
\newcommand{\X}{$X$}
\newcommand{\Xfac}{$X$ factor}
\newcommand{\Xnu}{$X_{\nu}$} 
\newcommand{\Xgal}{$X_{\rm Gal}$}
\newcommand{\siggas}{$\Sigma_{\rm gas}$}

\newcommand{\NCO}{$N_{\rm CO}$}
\newcommand{\NHt}{$N_{\rm H_2}$}
\newcommand{\nHt}{$n_{\rm H_2}$}
\newcommand{\nCO}{$n_{\rm CO}$}
\newcommand{\fCO}{$f_{\rm CO}$}
\newcommand{\NHtnu}{$N_{\rm H_2, \nu}$}
\newcommand{\W}{$W$}
\newcommand{\Ht}{H$_2$}
\newcommand{\deltav}{$\Delta v$}
\newcommand{\vlos}{$v_{los}$}
\newcommand{\Tb}{$T_B$}
\newcommand{\Tdv}{$T_B \Delta v$}
\newcommand{\MtoL}{$M_{VT}/L_{CO}$}
\newcommand{\taunu}{$\tau_\nu$}
\newcommand{\vmt}{$v_{\rm {mtrb}}$}

\newcommand{\vrms}{$v_{\rm rms}$}

\newcommand {\apgt} {\ {\raise-.5ex\hbox{$\buildrel>\over\sim$}}\ }
\newcommand {\aplt} {\ {\raise-.5ex\hbox{$\buildrel<\over\sim$}}\ }

\title[Modeling CO Emission from MCs: The \Xfac]{Modeling CO Emission:
  II. The Physical Characteristics that Determine the \Xfac\ in
  Galactic Molecular Clouds}
\author[R. Shetty et al.]{Rahul Shetty$^{1}$, Simon C. Glover$^{1}$, Cornelis P. Dullemond$^{1,2}$, Eve C. Ostriker$^{3}$, \and Andrew I. Harris$^{3}$, Ralf S. Klessen$^{1}$\\ \\
$^{1}$Zentrum f\"ur Astronomie der Universit\"at Heidelberg, Institut f\"ur Theoretische Astrophysik, Albert-Ueberle-Str. 2, 69120 Heidelberg, Germany \\
$^{2}$ Max Planck Institut f\"ur Astronomie, K\"onigstuhl 17, 69117 Heidelberg, Germany\\
$^{3}$ Department of Astronomy, University of Maryland, College Park, MD, 20742, USA }

\begin{document}

\date{Accepted 2011 April 18. Received 2011 April 15; in original form 2011 February 24}

\pagerange{\pageref{firstpage}--\pageref{lastpage}} \pubyear{2010}

\maketitle

\label{firstpage}

\begin{abstract}

We investigate how the \Xfac, the ratio of the molecular hydrogen
column density (\NHt) to velocity-integrated CO intensity ($W$), is
determined by the physical properties of gas in model molecular clouds
(MCs).  The synthetic MCs are results of magneto-hydrodynamic
simulations, including a treatment of chemistry.  We perform radiative
transfer calculations to determine the emergent CO intensity, using
the large velocity gradient approximation for estimating the CO
population levels.  In order to understand why observations generally
find cloud-average values of \X = \Xgal $\sim2\times10^{20}$ \Xunits,
we focus on a model representing a typical Milky Way MC.  Using
globally integrated \NHt and $W$ reproduces the limited range in
\X\ found in observations and a mean value \X = \Xgal
$=2.2\times10^{20}$ \Xunits.  However, we show that when considering
limited velocity intervals, \X\ can take on a much larger range of
values due to CO line saturation.  Thus, the \Xfac\ strongly depends
on both the range in gas velocities, as well as the volume densities.
The temperature variations within individual MCs do not strongly
affect \X, as dense gas contributes most to setting the \Xfac.  For
fixed velocity and density structure, gas with higher temperatures $T$
has higher $W$, yielding $ X \propto T^{-1/2}$ for $T\sim 20-100$ K.
We demonstrate that the linewidth-size scaling relationship does not
influence the \Xfac\ $-$ only the {\it range} in velocities is
important.  Clouds with larger linewidths $\sigma$, regardless of the
linewidth-size relationship, have a higher $W$, corresponding to a
lower value of \X, scaling roughly as $X \propto \sigma^{-1/2}$.  The
``mist'' model, often invoked to explain a constant \Xgal\ consisting
of optically thick cloudlets with well-separated velocities, does not
accurately reflect the conditions in a turbulent molecular cloud.  We
propose that the observed cloud-average values of \X$\sim$\Xgal\ is
simply a result of the limited range in \NHt, temperatures, and
velocities found in Galactic MCs $-$ a nearly constant value of
\X\ therefore does not require any linewidth-size relationship, or
that MCs are virialized objects.  Since gas properties likely differ
(albeit even slightly) from cloud to cloud, masses derived through a
standard value of the \Xfac\ should only be considered as a rough
first estimate.  For temperatures $T\sim 10-20$ K, velocity
dispersions $\sigma \sim 1-6$ \kms, and $N_{H_2} \sim 2-20 \times
10^{21}$ \cmsq, we find cloud-averaged values \X$\sim 2-4
\times10^{20}$ \Xunits\ for Solar-metallicity models. 

\end{abstract}

\begin{keywords}
ISM:clouds -- ISM: lines and bands -- ISM: molecules -- ISM: structure -- line:profiles -- stars:formation
\end{keywords}

\section{Introduction\label{introsec}}
Carbon monoxide (CO), the second most abundant molecule in the
interstellar medium (ISM), has now been observed for over thirty years
to investigate the physical characteristics of the ISM.  The lowest
rotational levels are easily excited through collisions with molecular
hydrogen (\Ht), by far the primary component of molecular gas in the
ISM.  Since \Ht\ is difficult to detect directly, and the $^{12}$CO
J=1-0 line occurs at a frequency (115.27 GHz) that is readily
observable from Earth, CO observations are well suited for probing the
conditions of the molecular component of the ISM.

Accurately measuring the masses and velocities of gas within molecular
clouds (MCs) is of primary importance for understanding star
formation.  CO observations of Galactic and extra-galactic MCs have
provided a wealth of information about these properties, allowing for
detailed modeling of the star formation process.  However, uncertainty
remains about exactly how to convert observed CO emission into
fundamental physical properties of the dominant molecular component of
the ISM.

The low rotational transition lines of CO are known to be optically
thick, and therefore a considerable fraction of the emission from high
density regions must be self-absorbed.  Nevertheless, a strong
correlation is found between the CO intensity and the \Ht\ column
density \NHt.  A number of methods are employed to measure \NHt.
These include mass determinations using observations of $^{13}$CO,
which has lower optical depth than $^{12}$CO and so may be capable of
tracing much of the MC gas \citep[e.g.][]{Dickman78}.  Alternatively,
if MCs are in virial equilibrium, the $^{12}$CO linewidth may be used
to estimate the virial mass \citep[e.g.][]{Larson81,Solomonetal87}.
Independent mass measurements not involving CO include observations of
$\gamma$-rays, which are produced when cosmic rays interact with the
ISM.  The molecular content is deduced when HI observations provide
information about the amount of atomic material along the
line-of-sight \citep[][]{Strongetal88}.  Dust-based observations in
the infrared may also provide indirect gas mass estimates, using
appropriate dust-to-gas ratios \citep[e.g.][]{Dameetal01,
  Pinedaetal08, Leroyetal09}.

The correlation between $^{12}$CO (J=1-0)\footnote{We will hereafter
  refer to $^{12}$CO (J=1-0) simply as ``CO.''} intensity \W\ and
\NHt\ is expressed as
\begin{equation}
X=\frac{N_{\rm H_2}}{W} \, ({\rm cm}^{-2}\,{\rm K}^{-1} \, {\rm km}^{-1} \, {\rm s}).
\label{Xfaceqn}
\end{equation}
In the Milky Way, observational analyses generally find this ``\Xfac''
to be nearly constant and $\sim$few$\times10^{20}$ \Xunits, hereafter
\Xgal, for both MCs \citep[see e.g.][and references
  therein]{Solomonetal87, Young&Scoville91} and the lower density
diffuse ISM \citep[e.g.][]{Polketal88, Lisztetal10}.  In regions with
very high molecular densities, such as ultraluminous infrared galaxies
(ULIRGs), \X\ is found to be $\sim$1-5 times lower than \Xgal.  There
is believed to be only a small range in the \Xfac\ among ULIRGs
\citep[see][and references therein]{SolomonVandenBout05}.  As a result
suitable \Xfac s, depending on environment, are commonly used for
directly estimating the \NHt\ (or gas mass) from CO observations.

A key assumption in using an \Xfac\ to derive gas masses is that the
CO line is an approximately linear tracer of the bulk of the MC gas.
Since the CO line is optically thick, however, it is not obvious why a
linear relationship should hold.  Extending the analysis of
\citet{Dickmanetal86}, \citet{Solomonetal87} advanced the ``mist''
model in which a MC is composed of optically thick cloudlets that have
well separated velocities, such that the amount of self-absorption in
the MC is small, and \W\ is proportional to the total number of
cloudlets along the line-of-sight, hereafter LoS.  The assumptions
implicit to this ``mist'' model have not, however, been tested with
radiative transfer modeling in realistic MC models.

Despite the correlations between \W\ and \NHt\ described above, there
are in fact signs that no universal \Xfac\ is applicable to all
sources.  First, since the CO line is optically thick, observations of
Galactic MCs show that beyond a threshold column density,
\W\ saturates so that the CO line no longer traces gas mass
\citep{Lombardietal06,Pinedaetal08}.  Observations of low metallicity
systems such as the SMC suggest \X$\gg$\Xgal\ \citep{Israeletal86,
  Israel97, Bosellietal97, Bosellietal02, Leroyetal09, Leroyetal11}.
Theoretically, low metallicity systems are expected to have lower CO
abundances, which lead to lower CO intensities and thereby larger
\Xfac s \citep{Maloney&Black88, Wolfireetal93, Israel97, Shettyetal11}
provided that the other properties of the MCs (e.g. mass, size) do not
vary greatly with metallicity \citep{Glover&MacLow11}.  Further, using
independent dust-based measures of \NHt\ leads to \Xfac\ estimates
that differ from \X\ computed through the ``virialized'' cloud
assumption \citep[][]{Bolattoetal08, Leroyetal07, Leroyetal09}.  This
discrepancy is taken as evidence that there are large reservoirs of
molecular gas untraced by CO \citep{Grenieretal05, Planck11}.  One
explanation for such a situation is that in the outer regions of MCs,
CO cannot form efficiently due to poor self-shielding
\citep{Wolfireetal10}.

The discrepancies and range in the \Xfac\ estimates need to be
understood in order to accurately interpret CO observations.  One
important issue related to the \Xfac\ is the dynamical state of MCs.
Two well known scaling relationships for MCs have been established in
large part due to CO observations: the mass-size and linewidth-size
relationships \citep{Larson81}, often referred to colloquially as
``Larson's Laws.''  Assuming constant \X, the MC masses $M$ are found
to be related to the projected size $R$ through a power law with
approximate index 2, $M \propto R^2$.  Observed linewidths for MCs
follow a power law relationship with projected size $\sigma \propto
R^{1/2}$.  Taken together, the interpretation of these relationships
is that clouds are (approximately) in virial equilibrium, or
``virialized,'' so $\sigma^2 \approx GM/R$ \citep{Larson81,
  Dickmanetal86, Solomonetal87, Myers&Goodman88}.  Alternatively, if
MCs are virialized such that $N_{\rm H_2} \sim M/(\mu R^2) \sim
\sigma^2/(\mu G R)$, then $X \sim \sigma^2/(\mu G R W)$.  If
$\sigma^2/R$ and $W$ vary little over the population of MCs, then
\X\ would have a uniform value.

An open question is whether the ``mist'' model is applicable to
turbulent media, as turbulence is now considered a dominant factor
controlling the dynamics of MCs \citep[e.g.][and references
  therein]{MacLow&Klessen04, McKee&Ostriker07}.  In the first
publication in this series (\citealt{Shettyetal11}, hereafter Paper
I), we investigated how well CO can trace the underlying molecular gas
using radiative transfer calculations on magnetohydrodynamic (MHD)
models of MCs, which include a treatment of chemistry.  We focused on
velocity integrated CO intensity, and compared probability
distribution functions (PDFs) of \W, \NHt, and CO column density \NCO.
We also assessed the \Xfac, again only considering velocity integrated
intensities.  We showed that even though \X\ may vary between
different LoSs through a given MC, the cloud-averaged intensity
produces \X$\approx$\Xgal\ within a factor of 3, from various models
with different metallicities $Z/Z_\odot =0.3-1$ and densities
$n_0=100-300$\cmt.  In this work, we focus primarily on the Milky Way
MC model from Paper I, and investigate the properties that affect the
derived \Xfac.  Using radiative transfer calculations of turbulent
chemo-MHD models, we perform a rigorous theoretical investigation of
the qualitative models proposed to explain the observed \Xfac.  We
directly modify the cloud characteristics, such as temperature,
density, and velocity, and recompute the \Xfac\ to understand its
dependence on those parameters.  Two of our goals are to understand
why the \Xfac\ is roughly constant for a range of systems, including
Milky Way field GMCs, and to assess whether the ``mist'' model is
applicable to turbulent MCs.

This paper is organized as follows.  In the next section, we provide
an overview of the estimated \Xfac\ in various environments.  We also
discuss the results from Paper I in the context of observationally
derived values.  In Section \ref{modsec} we review our method of
modeling CO emission from turbulent MCs, and discuss some properties
of the main Milky Way MC model.  In Section \ref{ressec}, after
discussing how the \Xfac\ is measured, we compare how \X\ depends on
the dynamic, chemical, and thermal structure of the model MC.  We
investigate in detail the dependence of \X\ on various cloud
characteristics, such as temperature, \Ht\ and CO densities, as well
as the velocities.  In Section \ref{discsec} we compare our results to
previous observational and theoretical efforts, and offer a new
explanation for observed trends.  Section \ref{sumsec} summarizes our
interpretation of the observed \X$\approx$\Xgal and our conclusions
regarding the parameter dependences of \X.

\section{Overview of the X factor}\label{motivationsec}

\begin{figure*}
\includegraphics[width=100mm]{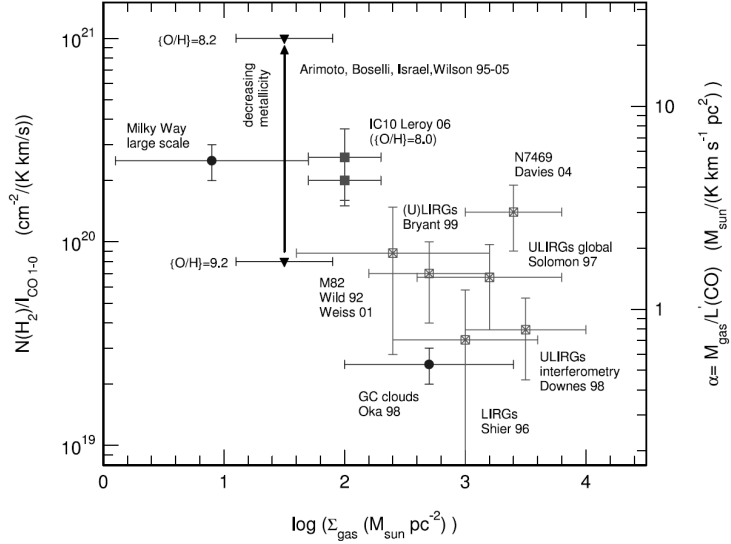}
\caption{Compilation of estimated \X\ factors from a range of systems,
  shown as a function of surface density.  Figure reproduced from
  \citet{Tacconietal08}.}
\label{Xenvirons}
\end{figure*}

As alluded to in the previous section, the \Xfac\ is found to be
nearly constant when considering a specific class of sources, such as
Galactic MCs, or a different value for ULIRGs.  Yet, analysis of the
full spectrum of molecular environments generally portray a trend of
decreasing \Xfac\ with increasing molecular surface density.  Figure
\ref{Xenvirons} shows a compilation of observationally inferred \Xfac
s from various systems \citep{Tacconietal08}.

\begin{figure*}
\includegraphics[width=100mm]{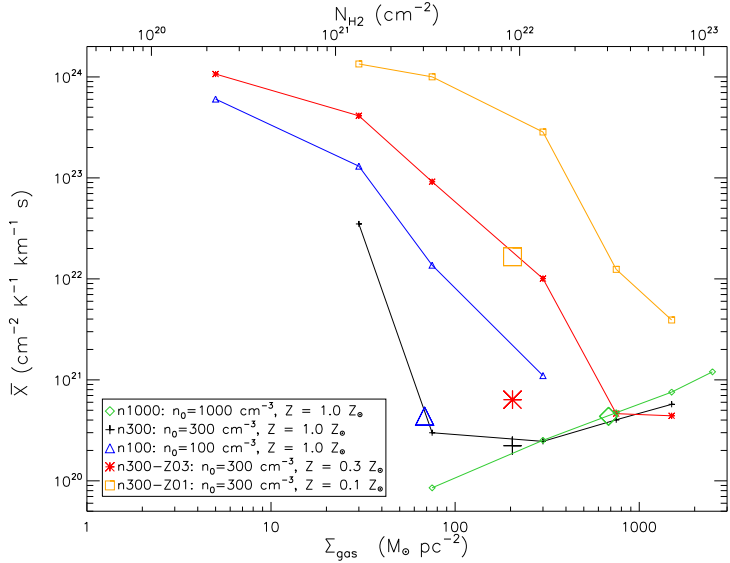}
\caption{Mean \X\ factor in bins of gas surface density
  \siggas\ (bottom axis) or \NHt\ (top axis) for 5 models.  The
  \X\ factor is averaged in different \siggas\ bins.  The value of
  $\overline{X}$ is plotted on the midpoint value of \siggas\ of each
  bin.  Each model is identified by different colors and symbols (and
  labeled in the legend).  The large symbols shows the global
  (emission-weighted) mean \X\ factor and mean \siggas\ from each
  model.}
\label{meanX}
\end{figure*}

In Paper I, we analyzed the \Xfac\ in various models with different
metallicities $Z/$\zsun = 0.1 - 1 and densities $n_0=100-1000$ \cmt.
Figure \ref{meanX} shows the relation between the \Xfac\ and surface
density \siggas\ or \NHt\ from the various models discussed in Paper
I.\footnote{The conversion between \NHt\ and \siggas\ includes a
  factor of 1.4 to account for the mass of helium.}  The points show
the mean \Xfac\ averaged in bins of \siggas, and the large points show
the cloud-averaged \Xfac.  At intermediate densities
\siggas$\sim$50-200 \msunpc, the cloud-averaged \Xfac s for all but
the very low $Z$=0.1 model are $\sim 2-5 \times 10^{20}$
\Xunits\ $\approx$ \Xgal.

Both Figs. \ref{Xenvirons} and \ref{meanX} show that \X\ is larger
than \Xgal\ in low metallicity systems.  As explained in Paper I,
large \Xfac\ values at low metallicities and densities are primarily
due to the low \W, the integrated CO intensity, relative to \NHt
(Eqn. \ref{Xfaceqn}).  Systems with lower abundances of CO will of
course have lower \W.  The formation of CO is highly dependent on the
MC density, metallicity, and strength of the background UV radiation
field \citep[see also][]{Maloney&Black88, vanDishoeck&Black88,
  Gloveretal10, Glover&MacLow11}.  However, \Ht\ formation is not as
sensitive to these properties, due to its ability to effectively
self-shield.  Thus, the relative abundance of CO compared to \Ht\ can
vary significantly within a MC, leading to a wide range in the \Xfac,
even though the (emission-weighted) average of many different clouds
may all result in a value $\sim$\Xgal.

Note that the low surface density regimes in Figs. \ref{Xenvirons} and
\ref{meanX} are not directly comparable.  The low surface densities in
our numerical models correspond to diffuse regions in MCs with size
$\sim$10 pc.  In Figure \ref{Xenvirons}, the low surface density cases
correspond to observations with low beam filling factors.  These
regions can have sizes $\sim$ kpc (or larger).  Therefore, the objects
in the two figures may differ in size.

At large surface densities (\siggas \apgt 200 \msunpc), our turbulent
models show an increase in the \Xfac, whereas the observations suggest
a decreasing \Xfac.  The increase of \X\ with increasing surface
density in our models occurs because in this regime the CO line is
saturated, so that \W\ remains constant even as \NHt\ increases.

Our models only account for the effects of MHD, thermodynamics, and
chemical evolution.  The sources at high surface densities in
Fig. \ref{Xenvirons} are either the Milky Way center or galaxies
undergoing intense star formation activity (LIRGs and ULIRGs).  The
heating associated with star formation, as well as the large-scale
rotation and turbulence of the ISM in these galaxies, are not captured
by our models.  These processes likely contribute to setting the
\Xfac\ in such environments, and may be responsible for the observed
trends in Figure \ref{Xenvirons} at high $\Sigma_{\rm gas}$.

In our current investigation of the \Xfac, we aim to understand which
MC properties are responsible for \X$\sim$\Xgal\ in the 50 - 200
\msunpc\ range.  To carry out our analysis, we perform simple
experiments by manually changing a number of physical parameters of
our models.  We then recompute the \X\ factor, and assess which of the
modified parameters most affect the resulting value of \X.

\subsection{Definition of the \Xfac}
We motivate our choice of the parameters to be modified by the
definition of the \Xfac.  When the CO intensity is expressed in units
of the Rayleigh-Jeans ``brightness temperature,'' $T_B$ then
\begin{equation}
W=\int T_B dv \,\, ({\rm K \, km \, s^{-1}}). 
\label{Weqn}
\end{equation}
The \Ht\ column density is simply the volume density integrated over
the LoS $ds$:
\begin{equation}
N_{\rm H_2}=\int n_{H_2} ds
\label{Nh2eqn}
\end{equation}
Taking Equations \ref{Xfaceqn}-\ref{Nh2eqn} together, 
\begin{equation}
X=\frac{\int n_{H_2} ds}{\int T_B dv}.
\label{Xfaceqnall}
\end{equation}
This indicates that the \Xfac\ is explicitly dependent on three
quantities: the column density of \Ht, the peak CO intensity, and the
range in velocities.  Due to the coupling between hydrodynamics,
thermodynamics, and chemistry, $T_B$ is also dependent on the velocity
and density (as well as the kinetic temperature).  We aim to
understand the relative contribution of each of these three properties
of the MC.  After assessing the \Xfac\ from the original Milky Way MC
model, we alter one of these properties at a time, while keeping the
others fixed, and recompute the \Xfac.  In this manner, we can
identify the most important cloud properties responsible for setting
the \Xfac.

\section{Modeling Method}\label{modsec}
\subsection{Numerical magnetohydrodynamics, chemistry, and radiative transfer}

To investigate how MC characteristics affect the \Xfac, we analyze
magnetohydrodynamic (MHD) models of molecular clouds that include a
time-dependent treatment of chemistry.  We perform radiative transfer
calculations on these numerical models, in order to solve for the CO
level populations and compute the emergent CO intensity.  The ratio of
the \Ht\ column density to the emergent CO intensity then gives the
\Xfac\ (Eqn. \ref{Xfaceqn}).

The MHD grid-based models follow the evolution of an initially fully
atomic medium with constant density in a $(20\, \pc)^3$ periodic box.
Thermodynamics is coupled with chemistry to follow the formation and
destruction of 32 chemical species, including \Ht\ and CO, through 218
chemical reactions.  Emission from CO and C$^+$ are the primary cooling
mechanisms in the dense and diffuse regions, respectively.
Additionally, a constant background UV radiation field is included,
which can photodissociate molecules in regions with insufficient
shielding.  The photoelectric effect is responsible for most of the
heating in the diffuse regions, and in more dense regions heating is
primarily due to cosmic ray interactions.  Turbulence is driven on
large scales (with wavenumbers $1 \le k \le 2$) to maintain an
approximately constant root mean square (3D) velocity \vrms.

For a thorough description of the modeling method, we refer the reader
to \citet{Glover&MacLow07I,Glover&MacLow07II} and
\citet{Gloveretal10}.  In this work, we primarily consider the
fiducial model chosen to match typical Milky Way MC conditions, which
has an initial hydrogen nuclei density $n_0$=300 \cmt, a metallicity
$Z/$\zsun = 1, a background UV radiation field\footnote{This value is
  1.7$G_0$ as determined by \citet{Draine78}, where $G_0$ is the
  \citet{Habing68} field.}  2.7$\times 10^{-3}$ erg cm$^{-2}$
s$^{-1}$, and a time averaged turbulent velocity \vrms=5 \kms.  The
magnetic field is initially oriented along the $\hat{z}$-axis, with
magnitude 1.95 $\mu G$.

We use the radiative transfer code
RADMC-3D\footnote{www.ita.uni-heidelberg.de/$\sim$dullemond/software/radmc-3d/}
to calculate the emergent CO intensity.  The level populations in each
zone are calculated through the Sobolev approximation
\citep{Sobolev57}, which uses velocity gradients across the faces of
each grid zone to estimate photon escape probabilities.  We employ the
Einstein and collisional rate coefficients estimated by
\citet{Yangetal10} and provided in the LAMDA database
\citep{Schoieretal05}.  A full description of the radiative transfer
calculation is provided in Paper I.

The radiative transfer calculations provide the CO intensities at each
LoS position ($x$,$y$) at a given frequency $\nu$ (or velocity) bin,
$I_\nu(x,y)$.  Besides the viewing geometry and the spectral
resolution of the synthetic observation, the only other user defined
parameter required to perform the calculation is the microturbulent
velocity \vmt\ (see Eqns. 2-7 in Paper I).  For our fiducial model, we
use \vmt = 0.5 \kms.  In the Appendix, we demonstrate that \W\ or
\X\ does not strongly depend on this choice, for \vmt $\in 0.25 -
0.75$ \kms.

Since we are interested in emission from all the gas in the simulation
volume, the spectral channels span the full range in gas velocities.
The position-position-velocity (PPV) cube provided by the radiative
transfer calculations has a spatial resolution of 0.08 pc or 0.16 pc,
corresponding to the extent of the zones in the 256$^3$ or 128$^3$
simulations, respectively, and a spectral resolution of 0.06 \kms.

\subsection{The fiducial Milky Way GMC model}\label{fidsec}

\begin{figure*}
\includegraphics[width=180mm]{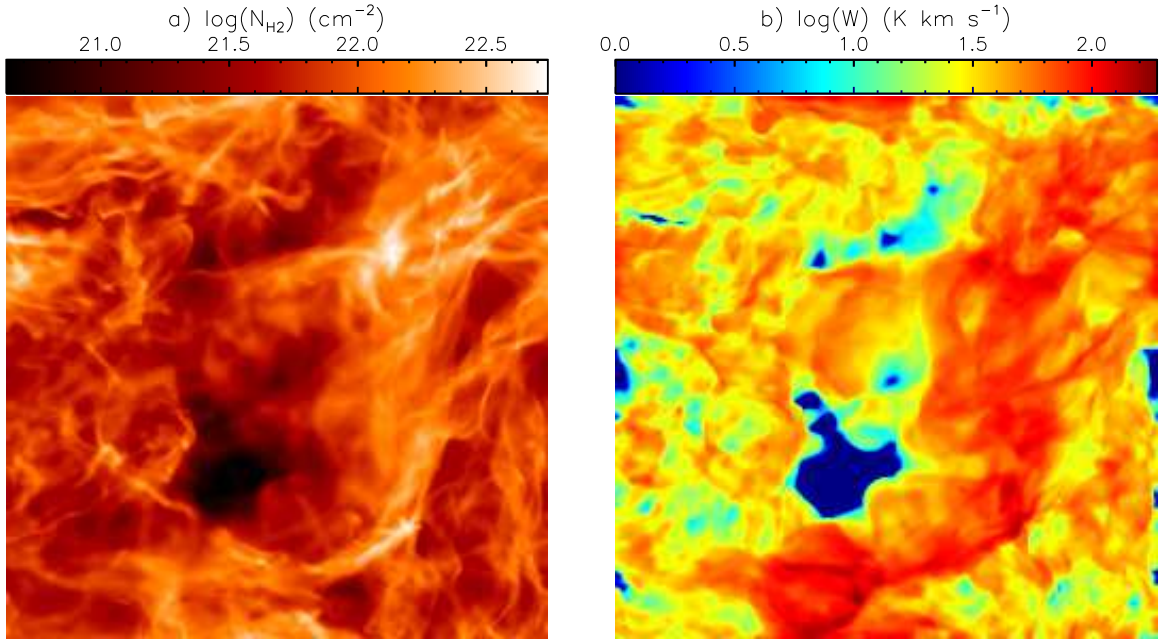}
\caption{a) Column density \NHt\ and b) integrated CO intensity of the
  Milky Way model MC.}
\label{fidim}
\end{figure*}

Figure \ref{fidim} shows \NHt\ and $W$ of the fiducial Milky Way model
MC, n300.  Though the overall morphology of \NHt\ is evident in the CO
image, there are some stark differences.  Most notably, the brightest
regions in the CO map (near the bottom) do not correspond to the
region with the highest column density (near the top right).  This
discrepancy arises due to the high optical depth in the CO line.  The
differences between the observed emission and the underlying gaseous
properties of these and other models are described in more detail in
Paper I.

Table \ref{simprops} lists the mass and volume weighted \Ht\ number
densities, temperatures, and LoS rms velocities\footnote{Since the
  viewing angle is along the $\hat{z}$-axis, $v_{los}= v_z$. Due to
  large-scale stochastic variations, the dispersion in $v_z$ is not
  exactly equal to $v_{\rm rms}/\sqrt{3}$.} $\langle
v_{los}^2\rangle^{1/2} = \sigma_{v,los}$, along with the initial
density $n_0$ and box size $L$ of model n300.  Mass and volume
weighted quantities are defined as $\langle f \rangle_{mass} = \sum f
\rho dV / \sum \rho dV$ and $\langle f\rangle_{vol} = \sum f dV / \sum
dV$, respectively.\footnote{$\langle n_{H_2}\rangle _{vol}$ is not
  exactly 150 \cmt\ because a small fraction of hydrogen remains
  atomic \citep{Glover&MacLow11}.}  The last row of Table
\ref{simprops} shows a reference value of the \Xfac: $\langle X\rangle
_{ref} \equiv \langle n_{H_2}\rangle_{vol} L / [\langle
  T\rangle_{mass} \sigma_{v,los} ]$.

\begin{table}
 \centering
  \caption{Characteristics of Standard ``Milky Way GMC'' Model (n300)}
  \begin{tabular}{cc}
  \hline
  \hline
Property & Value  \\
 \hline
 Box Size $L$ & 20.0 pc \\
 Initial Atomic Density $n_0$ & 300.0 cm$^{-3}$ \\
\\
 $\langle n_{H_2}\rangle _{mass}$ & 1098.0 cm$^{-3}$ \\
 $\langle n_{H_2}\rangle _{vol}$ & 145.9 cm$^{-3}$ \\
\\
 $\langle T\rangle _{mass}$ & 19.8 K \\
 $\langle T\rangle _{vol}$ & 51.7 K \\
\\

$\sigma_{v,los, mass} = [\langle v_{los}^2\rangle _{mass}]^\frac{1}{2}$ & 2.4 \kms  \\
$\sigma_{v,los, vol} = [\langle v_{los}^2\rangle _{vol}]^\frac{1}{2}$ & 2.4 \kms \\
\\
 $\langle X\rangle _{ref}$ & 1.9 $\times 10^{20}$ \Xunits  \\
\hline
\end{tabular}
 \label{simprops}
\end{table} 

As discussed in Paper I and \citet{Gloveretal10}, the internal
properties of the fiducial Milky Way cloud model take on a range on
values.  For example the CO abundance relative to \Ht, \fCO, can vary
from $\sim 10^{-9}$ to $10^{-4}$.  Similarly, the temperature,
density, and velocity also take on a wide range of values.  Throughout
our analysis, besides considering CO emission from the original model,
we also consider models for which some relevant physical
characteristics are modified, such as those with constant temperature
or CO abundances.  In this manner, we can quantitatively assess the
effect of the various physical characteristics on the emergent CO
intensity, and ultimately on the \Xfac.

\section{Results}\label{ressec}

\subsection{Measuring the \Xfac\ } \label{Xmeasure}

\begin{figure}
\includegraphics[width=90mm]{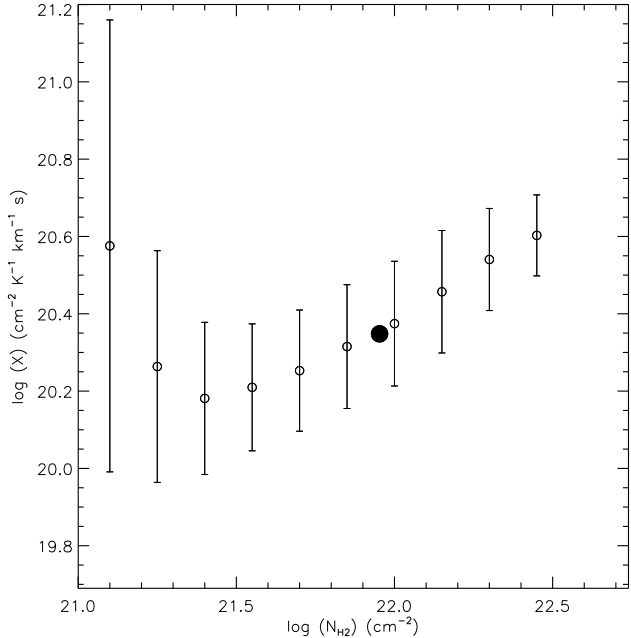}
\caption{\Xfac\ from the Milky Way model molecular cloud.  The open
  circles and error bars indicate the mean and standard deviation of
  the \X\ factor in bins of \NHt.  The solid circle shows emission
  weighted mean \Xfac\ of the whole model ($\langle X\rangle
  $=2.2$\times 10^{20}$ \Xunits) at the mean column density
  ($\overline N_{H_2} =9.0 \times 10^{21}$ \cmsq, corresponding to
  $\overline \Sigma_{\rm gas}=$ 202 \msunpc). }
\label{XvsN}
\end{figure}

We begin by calculating the \Xfac\ through its traditional definition
given by Equation \ref{Xfaceqn}.  Figure \ref{XvsN} shows the mean
\Xfac\ in bins of \NHt\ as a function of \NHt\ from all LoSs through
the fiducial model\footnote{Though the figures presented here only
  show the results from one viewing angle (along the $\hat{z}$-axis)
  we have verified that the results are not sensitive to any chosen
  viewing direction.}.  As explained in Paper I, at the highest
densities \W\ does not increase with increasing \NHt\ due to the
saturation of the CO line, resulting in \X$\propto$\NHt.
Nevertheless, the mean \Xfac\ only varies between (1.5 -4)$\times
10^{20}$ \Xunits\ for $\log$\NHt\ = 21-22.5.  Given this limited range
and the extent of the error bars, a constant value at its
emission-weighted mean value\footnote{We denote any weighted-mean
  value with $\langle ... \rangle$, whereas simple averages with an
  ``overline'' (e.g. $\overline{W}$.)} (shown by the large filled
circle in Fig. [\ref{XvsN}]) adequately describes the \Xfac\ for this
model.  This mean \Xfac\ value $\langle X \rangle $=2.2$\times
10^{20}$ is in good agreement with the reference value provided in
Table \ref{simprops}, $\langle X \rangle $=1.9$\times 10^{20}$.  We
now consider the terms \NHt\ and \W\ in detail, and their relationship
to the physical properties of the model.

Instead of integrating Equation \ref{Nh2eqn} over the whole cloud, one
could only use a limited range in velocity (or frequency, $\nu$).
\begin{equation}
N_{\rm H_2, \nu}={\int n_{H_2} dv}.
\label{3neqn}
\end{equation}
If \NHtnu\ is computed along the same LoS and in velocity bins that
match the spectral channels of the CO observation, then a column
density cube of \NHtnu\ can be constructed which has the same
configuration as the observed CO position-position-velocity (PPV)
cube.

We can now define the \Xfac\ at each PPV location, \Xnu, associated
with column density \NHtnu, intensity $T_B$ and adopted channel width
$\Delta v$ (in our case, 0.06 \kms):
\begin{equation}
X_{\nu} = \frac{N_{\rm H_2, \nu}}{{T_B \Delta v}}.
\label{3Xeqn}
\end{equation}

\begin{figure}
\includegraphics[width=90mm]{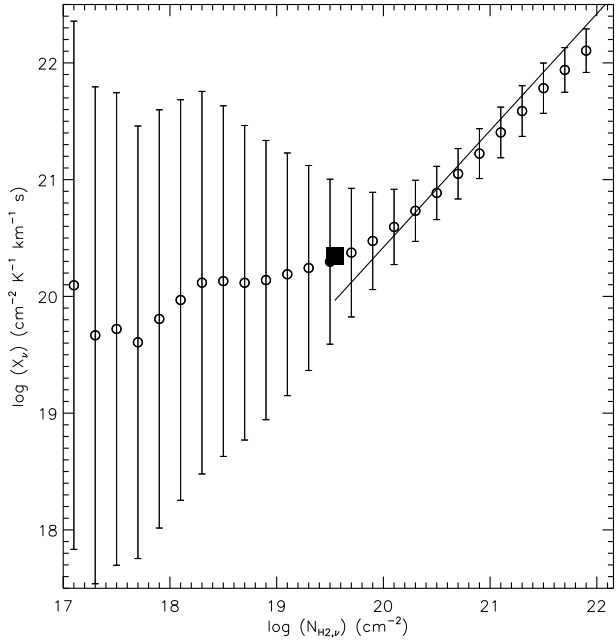}
\caption{\Xnu\ vs. \NHtnu\ from the Milky Way model molecular cloud.
  The mean and standard deviation of \Xnu\ in bins of \NHtnu\ are
  shown as circles with error bars.  The solid square shows emission
  weighted mean \Xnu\ of the whole model ($\langle X_\nu \rangle
  $=2.2$\times 10^{20}$ \Xunits) at the mean column density
  ($\overline{N_{\rm H_2, \nu}} =3.5 \times 10^{19}$ \cmsq).  Line
  shows \NHtnu / ($\overline{T_B} \Delta v$) at \NHtnu $>
  \overline{N_{\rm H_2, \nu}}$.}
\label{3DXvsN}
\end{figure} 

Figure \ref{3DXvsN} shows how this ``3D \Xfac'' depends on the column
density for the Milky Way model cloud.  Here, \NHtnu\ extends to much
lower values than the values of \NHt\ shown in Figure \ref{XvsN},
since the volume densities are integrated over a more limited region
(in velocity space).

The emission weighted mean ``3D \Xfac'' is equal to \Xfac\ computed in
its usual way, shown in Figure \ref{XvsN}.  However, at high
densities, \Xnu\ is significantly larger that \X, reaching values
\apgt$10^{22}$ \Xunits, whereas the 2D \Xfac\ is always \aplt
$10^{21}$ \Xunits.  From Figure \ref{3DXvsN}, it is clear that the CO
line is saturated at column densities (per frequency bin) above the
mean column density $\overline{N_{\rm H_2, \nu}} = 3.5 \times 10^{19}$
\cmsq, as indicated by the \NHtnu / ($\overline{T_B} \Delta v$) line,
where $\overline{T_B}$ is the mean brightness temperature in PPV
regions where \NHtnu $> \overline{N_{\rm H_2, \nu}}$.

\begin{figure}
\includegraphics[width=90mm]{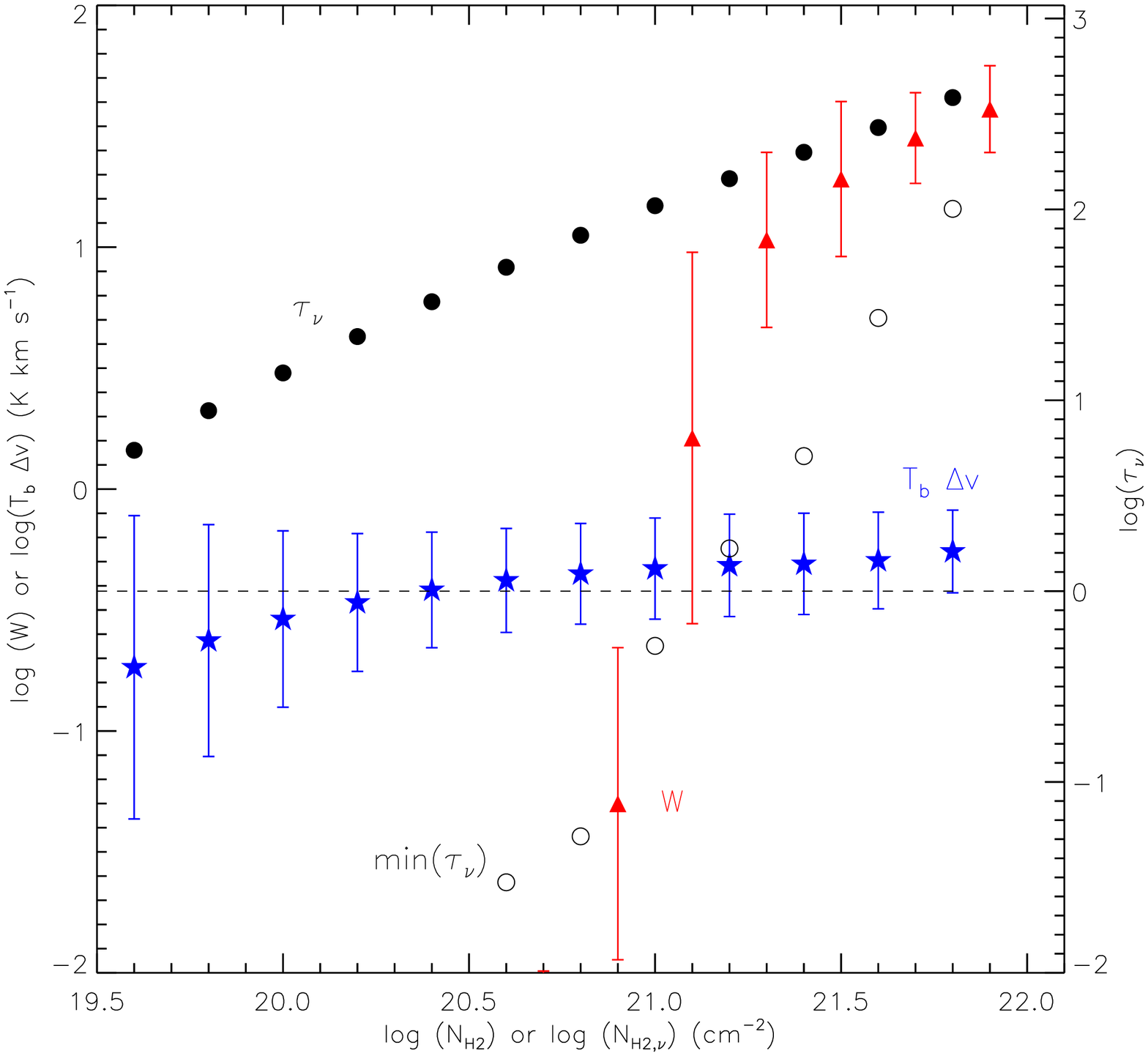}
\caption{Intensity (left ordinate) and optical depth (right ordinate)
  plotted against column densities.  Triangles show the mean
  integrated CO intensity \W\ as a function of total column density
  \NHt (2D).  Stars show the analogous relationship from (3D) PPV
  cubes, $T_B \Delta v$ vs \NHtnu.  Filled circles show the mean
  optical depth $\tau_\nu$, and the open circles show the minimum
  value of $\tau_\nu$ in the \NHtnu\ bins.  Dashed line corresponds to
  $\tau_\nu$=1. }
\label{WvsN}
\end{figure}

The discrepancy at high column densities between \X\ and \Xnu\ is due
entirely to the differences between \W\ and \Tdv.  On a given LoS,
\W\ is the summation of \Tdv\ through all velocities in the PPV cube.
Figure \ref{WvsN} shows these quantities as a function of \NHt\ or
\NHtnu, respectively.  Clearly, \W\ continues to increase beyond \NHt
\apgt $10^{21}$ \cmsq, whereas \Tdv\ is saturated.  Since $\Delta v$
is constant, this means that $T_B$ saturates at a value $\sim 13$
K.\footnote{This saturation brightness temperature is estimated from
  Figure \ref{WvsN}, which shows the mean values of \Tdv\ in bins of
  $N_{\rm H_2, \nu}$.  The maximum $T_B$ can reach values \apgt 30 K.}

Line saturation occurs when the optical depth at a given frequency
$\nu$, \taunu \apgt 1.  The optical depth along a LoS of
length $s$ is given by
\begin{equation}
\tau_\nu=\int \frac{h \nu}{4\pi} (n_1 B_{12} - n_2 B_{21}) \phi_\nu ds, 
\label{taueqn}
\end{equation}
where $h$, $n_i$, and $B_{ij}$ are the Planck constant, population
number density in level $i$, and the Einstein (stimulated
absorption/emission) coefficients.  The normalized line profile
function $\phi_\nu$ depends on the LoS and microturbulent velocities,
as well as the kinetic temperature (see Section 2.2 of Paper I for
more details).  Together, the density, temperature, and velocity
structure of the molecular cloud determines the optical depth, and
therefore where the line is saturated.

The mean and minimum value of \taunu, in bins of \NHtnu, are also
shown in Figure \ref{WvsN}, with the scale given on the right
ordinate.  Though \Tdv\ does not vary much in the range of the high
column densities shown in Figure \ref{WvsN}, a trend of increasing
\Tdv\ with column density is apparent until the \NHtnu $\approx
10^{20.5}$ \cmsq.  At column densities \NHtnu $\approx 10^{21}$ \cmsq,
all values of \taunu $> 1$, indicating complete saturation of \Tdv.

That the velocity-integrated brightness temperature \W\ still
increases beyond \NHt \apgt $10^{20.5}$ \cmsq, where the line is
saturated, indicates that numerous optically thick regions at
different $v$ contribute to \W.  This results in the saturation of the
integrated intensities occurring at higher column densities than the
saturation of the intensities in the PPV cube, as evident at the
highest \NHt\ in Figure \ref{WvsN}.

One consequence of the line saturation described above is that the
amount of gas that is untraced depends not only on the density, but
also on the velocity.  For instance, a ``cloudlet'' or parcel of dense
gas with similar velocity as another parcel along a LoS, but separated
in space, would not contribute extra flux to $T_B$, or the \W\ map.
On the other hand, a ``cloudlet'' with much different velocity along
the LoS would be traced, detected as higher $T_B$ at a different
location in the velocity axis of the spectrum, and thus contribute to
the integrated intensity $W$.  Consequently, LoSs with (dense) gas
spanning a larger range in velocities will have larger \W s.  This
concept of optically thick dense ``cloudlets'' is the basis of the
``mist'' model put forth by \citet{Solomonetal87} to explain the
uniformity in the \Xfac\ in Galactic GMCs.  We return to the
applicability of the ``mist'' model in Section 4.5.

\begin{figure}
\includegraphics[width=90mm]{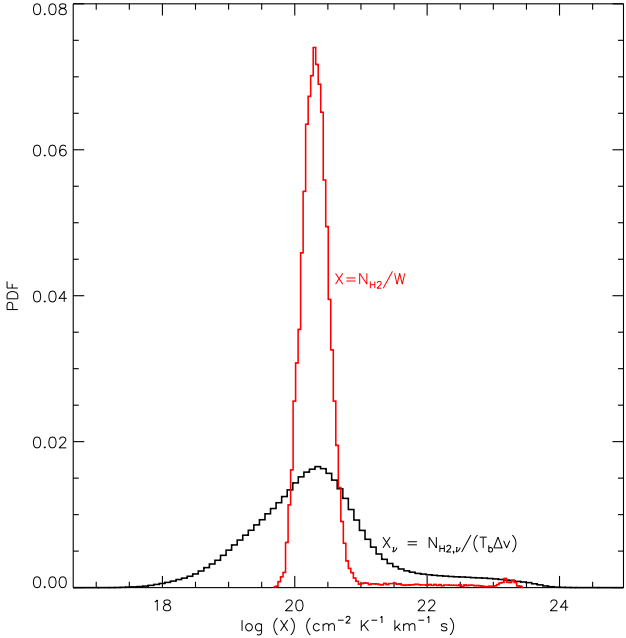}
\caption{PDFs of \Xfac s shown in Figures \ref{XvsN} and \ref{3DXvsN}. }
\label{Xhists}
\end{figure}

The differences in the relationships between \W\ and \Tdv\ with column
density (Fig. \ref{WvsN}) lead to the variations between \X\ and
\Xnu\ (Figs. \ref{XvsN} and \ref{3DXvsN}).  The (2D) \X\ and (3D)
\Xnu\ distributions are shown in Figure \ref{Xhists}.  Though the peak
value of the two distributions are similar, there is a much larger
distribution in \Xnu.  In considering synthetic observations with
different spectral resolution, we find that the range in \Xnu\ depends
on $\Delta v$, but that $\langle X_\nu \rangle $=2.2$\times 10^{20}$
\Xunits\ consistently.  As discussed, there are lower column densities
\NHtnu\ in a PPV cube compared to \NHt.  For such densities, the line
is often optically thin, and there is a range of possible emergent
intensities for a given column density (Fig. \ref{WvsN}).  This
results in a larger range in \Xnu\ compared to \X.

Taken together, the limited range in the 2D \Xfac\ shown in Figure
\ref{XvsN} and \ref{Xhists} only occurs when the CO intensities are
integrated over all velocities, and densities over the whole LoS.
Effectively, a limited range in the \Xfac\ only results when combining
the detailed velocity, as well as density, structure along the LoS.
This is indicative that the velocity range plays a crucial role in
setting the \Xfac.

Of course, integrating the observed brightness temperatures over a
limited range in velocity is not practical, as obtaining $N_{\rm H_2,
  \nu}$ is not readily feasible in observational data sets.
Nevertheless, such an analysis has underscored that the \Xfac\ depends
on the total velocity width of the cloud, as explicitly evident in
Equation \ref{Xfaceqnall}.  But, how does the detailed velocity
structure, such as its relationship with the size of emitting regions,
affect the \Xfac?  Further, what is the relative contribution of $\int
dv$ compared to the other terms $n_{H_2}$ and $T_B$ in Equation
\ref{Xfaceqnall}?  To address these questions, we investigate in
detail the role of each quantity in determining the nature of the CO
emission, and their relationship to intrinsic cloud properties.

\subsection{X factor dependence on temperature}\label{tempsec}

Due to the thermodynamics in the chemical-MHD model, the gas has a
range of temperatures.  As indicated in Table \ref{simprops}, the
model MC has a (volume-weighted) mean $T\approx$ 50 K, with a
dispersion $\sigma \approx$ 44 K.  This range in temperatures will
also result in a range in the \Xfac, since the observed brightness
temperature depends on the gas kinetic temperature (Eqn. \ref{Weqn}).

To test the sensitivity of the \Xfac\ to the temperature distribution,
we artificially set all temperatures in the model to a constant value,
and then perform the radiative transfer calculations.  The resulting
maps are then compared with the original CO map.  Any discrepancies
can be attributed to the differences in temperatures.

\begin{figure*}
\includegraphics[width=120mm]{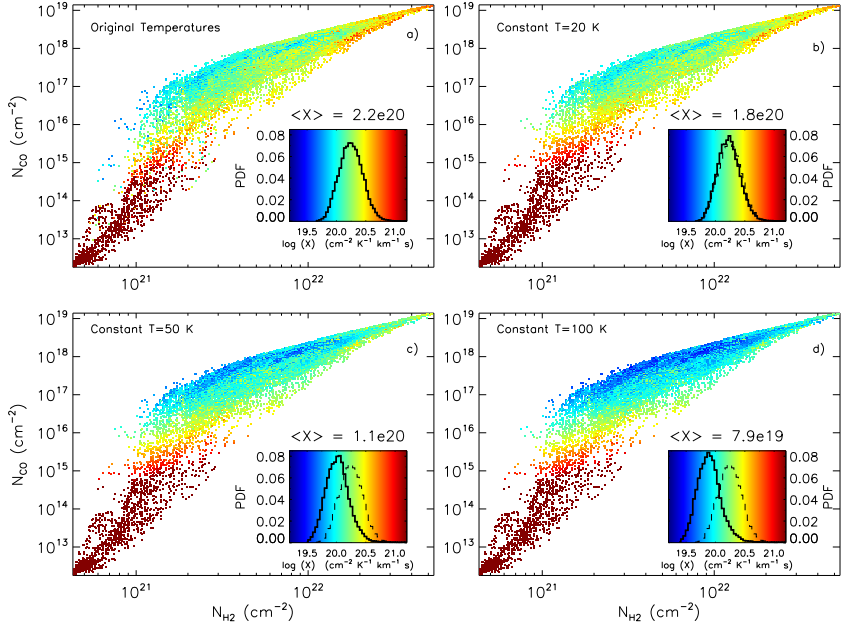} 
\caption{The variation of the \Xfac\ with gas temperature.  The
  position of each point shows the relationship between \NHt\ and
  \NCO.  The color of each point identifies the value of the \Xfac,
  computed through Eqn. \ref{Xfaceqn}, with the color scale and
  distribution given in the inset plots.  The emission-weighted mean
  value $\langle X\rangle $ is indicated in each plot.  a) Original
  model n300, b) model n300, but with constant T=20 K, c) constant
  T=50 K, and d) constant T=100.  In b), c), and d),
  \Xfac\ distribution from original simulation in a) is shown as thin
  dashed histogram.}
\label{tvar}
\end{figure*}

This is illustrated in Figure \ref{tvar}, where (a) shows the original
distribution of \X, and (b) - (d) show the \Xfac s for constant
temperatures of 25, 50, and 100 K.  The histograms from the model with
the original temperatures are provided in each panel as dashed lines.

The constant temperature $T$ = 20 K model, equal to the mass-weighted
temperature (see Table \ref{simprops}), provides an
\Xfac\ distribution that is very similar to the original model
(Figs. \ref{tvar}b).  This is expected since most of the CO is located
in dense regions where $T$ \aplt 20 K.  The relationship between \X,
\NHt, and \NCO\ is also rather comparable to the original
relationship.  Note that for both the original and constant $T$=20 K
models, the highest and lowest values of \NHt\ have large
\X. (cf. Fig. \ref{XvsN}).  The differences between the 20 K and 50 K
model suggests that the \Xfac\ is more sensitive to the mass-weighted
rather than the volume-weighted temperatures.

\begin{figure}
\includegraphics[width=90mm]{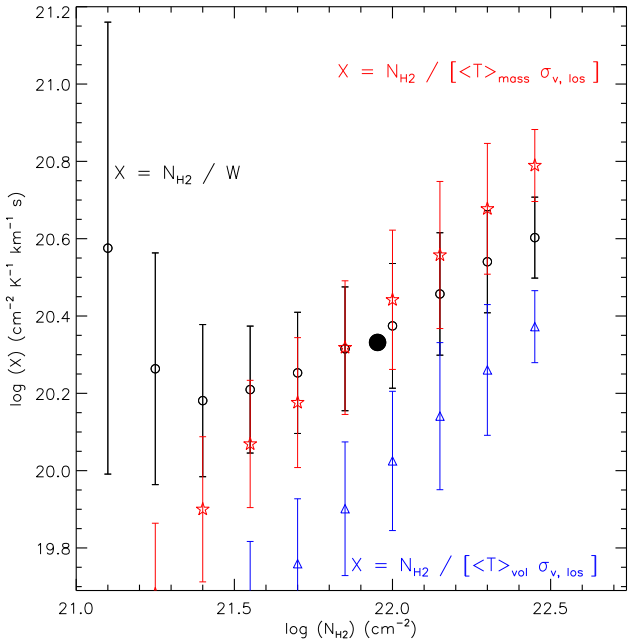}
\caption{ \Xfac\ (black circles) from the Milky Way model molecular
  cloud, as in Fig. \ref{XvsN}.  The emission-weighted mean \X\ at the
  mean column density is shown by the large solid circle.  Red stars
  show the the \Xfac\ computed by \NHt $/\langle T\rangle _{mass}
  \sigma_{v, los} $, and blue triangles show \NHt $/\langle T\rangle
  _{vol} \sigma_{v, los} $.  $\langle T\rangle_{mass}$ and $\langle
  T\rangle _{vol}$ are the global mass-weighted and volume-weighted
  temperatures, given in Table \ref{simprops}.  $\sigma_{v, los}$ is
  the mass-weighted rms velocity (which is equal to the
  volume-weighted rms velocity) along each LoS.}
\label{xvsnt}
\end{figure}

Figure \ref{xvsnt} shows the mean \X\ in bins of \NHt\ from the
original model (as in Fig. \ref{XvsN}), along with two reference
values obtained directly from the simulation.  The red stars show
\X\ computed using the mass-weighted LoS velocity and global
mass-weighted temperature.  The blue triangles show the corresponding
\X\ from the the volume-weighted quantities.  Along LoSs with column
densities lower than \aplt 10$^{21.8}$ \cmsq, the CO fraction
decreases, thereby decreasing W (see Paper I and Section
\ref{coabundsec}).  Thus, the simple relationships between \X, $T$,
and \vlos\ cannot reproduce the \Xfac\ at low column densities.  At
column densities \apgt 10$^{21.8}$ \cmsq, however, the \Xfac\ computed
using the mass-weighted temperature agrees fairly well with the
original \Xfac.  The \Xfac\ computed from the volume-weighted
temperature systematically underestimates the \Xfac.  These trends
indicate that the densest regions contribute most to setting \W, and
thereby the \Xfac.

As indicated in Figure \ref{tvar}, clouds with higher temperatures
produce lower \Xfac s, as would expected from
Eqn. \ref{Xfaceqn}-\ref{Weqn}.  From $\langle X\rangle = 1.8 \times
10^{20}, 1.1 \times 10^{20}$, and $7.9 \times \times
10^{19}$\Xunits\ at $T=20, 50$, and 100 K, respectively, we find an
approximate scaling $\langle X \rangle\propto T^{-0.5}$.  The range in
the \Xfac s in the modified-temperature models is, however, similar to
the original model.  This indicates that the range in temperatures in
model n300 is not responsible for the range in the \Xfac.  As a
result, it is either the range in densities or the velocities which
produce the distribution in the \Xfac.

We note that an increase in temperature to 100 K is likely accompanied
by other variations in the molecular cloud structure, such as a
decrease in \nHt\ and possibly \fCO\ if the high temperature is due to
a high star formation rate.  As such, models for which the
temperatures are manually scaled to high values are likely omitting
real environmental effects that would affect the \X\ - $T$
relationship.

\subsection{X factor dependence on density}

We turn our attention to the \Xfac\ dependence on density.  Any given
cloud will have a range in volume and column densities, in both
\Ht\ and CO.  If \W\ depends linearly on \NHt, then there would
certainly be a constant \Xfac\ along all LoSs.  However, as discussed
in Paper I, \W\ generally does not have a straightforward scaling with
the column density \NCO, especially at large \NCO\ due to the high
opacity of CO (see Section \ref{Xmeasure}).  Further, \NCO\ does not
directly trace \NHt.  This lack of correlation between \W\ and
\NHt\ results in a distribution of \X\ along different LoSs for a
given MC.  We focus on how the volume density and CO abundance affect
the \Xfac.

\subsubsection{X factor dependence on volume density}

To test whether the \Xfac\ is more sensitive to \nHt\ or to \NHt, we
consider a model with a higher volume density than model n300.  This
model (n1000), only differs from model n300 in its initial atomic
density of 1000 \cmt.  As described in Paper I, most of the carbon in
this model is incorporated into CO, and so there is a very good
correlation between \NCO\ and \NHt\ (see Fig. 5c in Paper I).

\begin{figure}
\includegraphics[width=90mm]{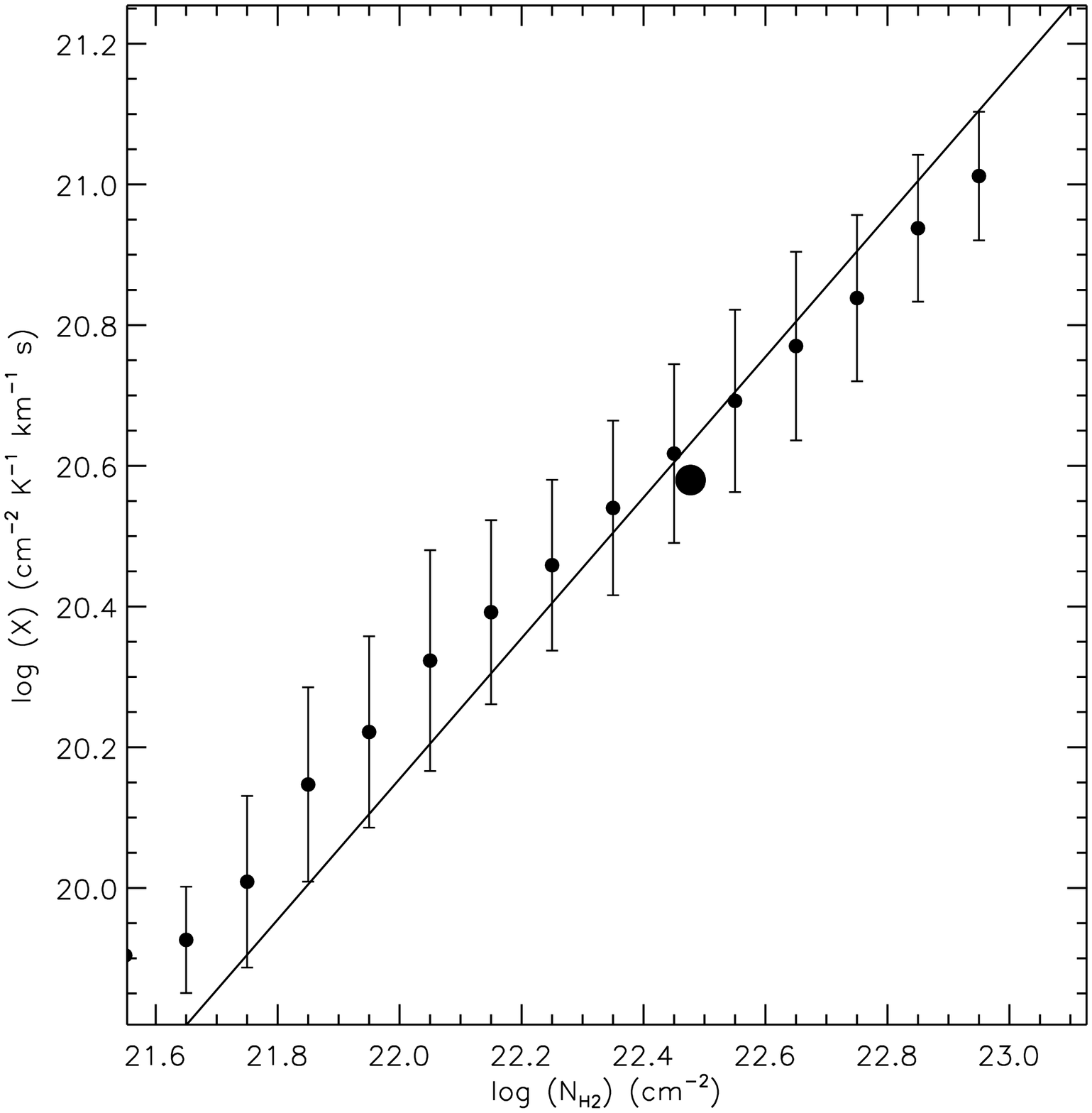}
\caption{\Xfac\ vs. \NHt\ from model n1000, which has box size (20
  pc)$^3$.  The line shows $X=N_{H_2}/\overline{W}$, with
  $\overline{W}$ = 70 \Kkms.  The large circle shows $\langle X\rangle
  $=3.8$\times 10^{20}$ \Xunits\ at $\overline N_{H_2} =3.0 \times
  10^{22}$ \cmsq, or $\overline \Sigma_{\rm gas}=$ 672 \msunpc.}
\label{Xn1000}
\end{figure}

As explained in Paper I, for this high density model the effect of
line saturation is clearly apparent in the \Xfac\ computed through
Equation \ref{Xfaceqn}.  Figure \ref{Xn1000} shows the relationship
between \X\ and \NHt\ for model n1000.  The line is not a fit, but
rather the relationship $X=N_{H_2}/\overline{W}$, where
$\overline{W}=$ 67 \Kkms\ is the mean integrated intensity from this
model.  We see that \X\ invariably increases with \NHt, as would be
expected if the line were saturated.

The \X $-$\NHt\ relationship for model n300 (Fig. \ref{XvsN}) is quite
different from the relationship from model n1000, due partly to the
larger maximum column density in model n1000 and lower minimum column
density in model n300.  We investigate whether the total column
density, which occurs explicitly in Equation \ref{Xfaceqn} is
responsible for the increase in \X\ at large \NHt in model n1000.  We
perform the radiative transfer calculations on a model which is
similar to n1000, but whose box length is decreased by a factor of
$\sim$3 to 6 pc.  In this n1000-L6 model, the lower column densities
allow the background UV radiation to penetrate further into the cloud,
leading to more CO photodissociation.  Nevertheless, this model should
have similar \Ht\ column densities to the n300 model, since \Ht\ can
self-shield more efficiently.  On the other hand, the volume densities
in n1000-L6 will be very similar to those in the n1000 model.

\begin{figure}
\includegraphics[width=90mm]{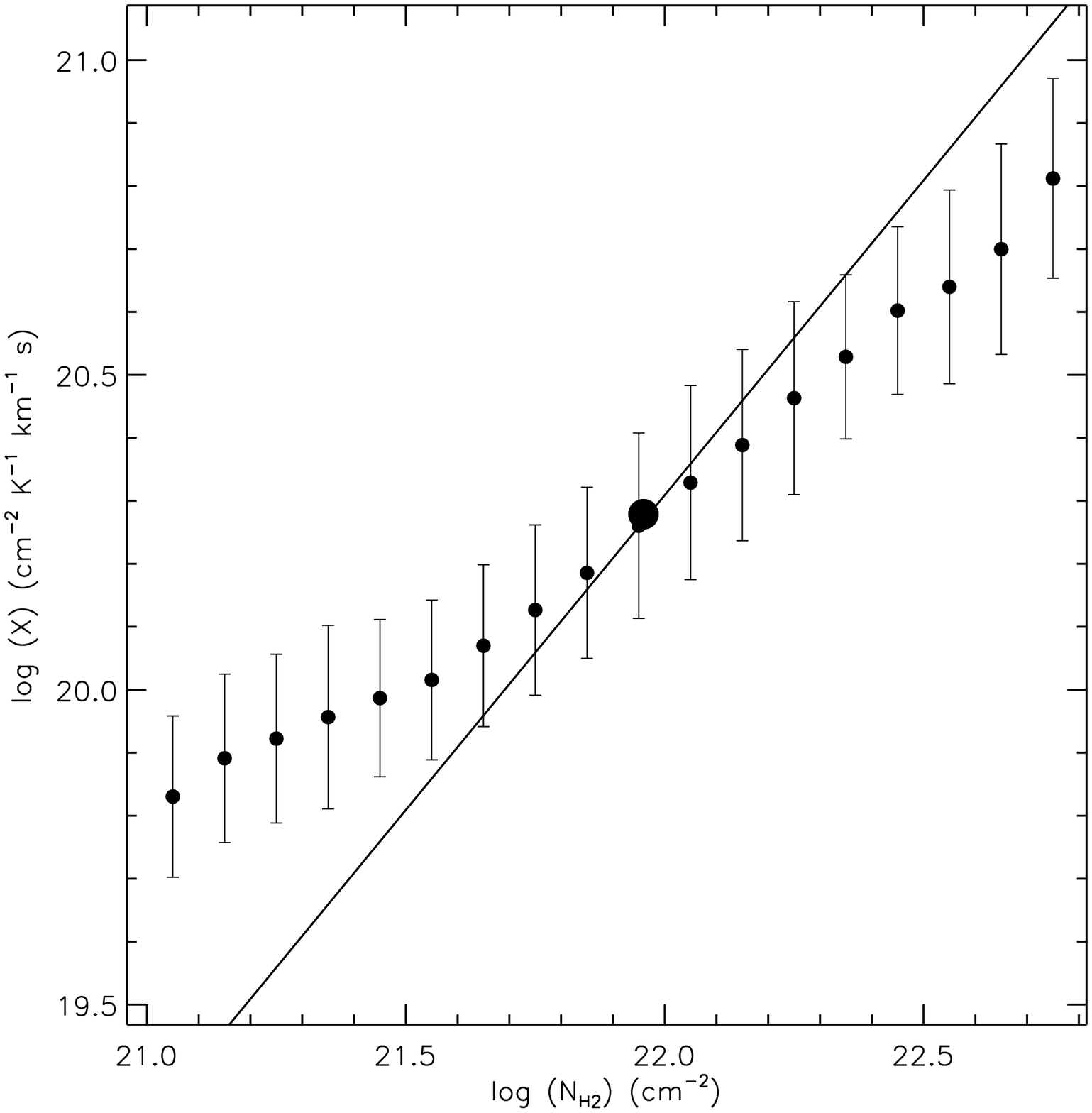} 
\caption{\Xfac\ vs. \NHt\ from model n1000-L6, which has initial
  density 1000 \cmt\ and box size (6 pc)$^3$.  Line shows
  $X=N_{H_2}/\overline{W}$, with $\overline{W}$ = 49 \Kkms.  The
  large circle shows $\langle X\rangle $=1.9$\times 10^{20}$
  \Xunits\ at $\overline N_{H_2} =9.1 \times 10^{21}$
  \cmsq\ corresponding to $\overline \Sigma_{\rm gas}=$ 204 \msunpc. }
\label{Xn1000split}
\end{figure}

Figure \ref{Xn1000split} shows the \X $-$\NHt\ relation for this
n1000-L6 model.  The line shows $X=N_{H_2}/\overline{W}$, with
$\overline{W}=49$ \Kkms\ for this model.  Here, $\langle X\rangle $ is
similar to the n300 model.  The \X $-$\NHt\ relation is not as well
reproduced by the $N_{H_2}/\overline{W}$ line as model n1000 in Figure
\ref{Xn1000}.  Nevertheless, the \Xfac\ clearly increases with \NHt.
Compared to model n300 (Fig. \ref{XvsN}), \X\ is similar for
\NHt\ $\sim 10^{22}-10^{22.5}$\cmsq, but there are significant
differences at lower \NHt, especially \aplt 10$^{21.5}$.  Thus, {\it
  the \Xfac, even when calculated using the total column density
  (Eqn. \ref{Xfaceqn}), depends on the volume density \nHt, rather
  than just the column density \NHt.}  Note, however, that when
averaged over all lines of sight, $\langle X\rangle$ decreases by only
15\% even though $n_0$ increases by a factor 3.3.

\subsubsection{X factor dependence on CO abundance}\label{coabundsec}

Besides containing a range in densities, there is also a range in CO
abundances at a given \Ht\ density, as is evident in the column
density relationships in Fig. \ref{tvar} (see also Paper I, and
\citealt{Gloveretal10}).  In order to investigate how this
distribution contributes to the \Xfac, we reset the CO abundance with
a constant \fCO = \nCO/\nHt\ ratio, and then perform the radiative
transfer calculations.

\begin{figure*}
\includegraphics[width=120mm]{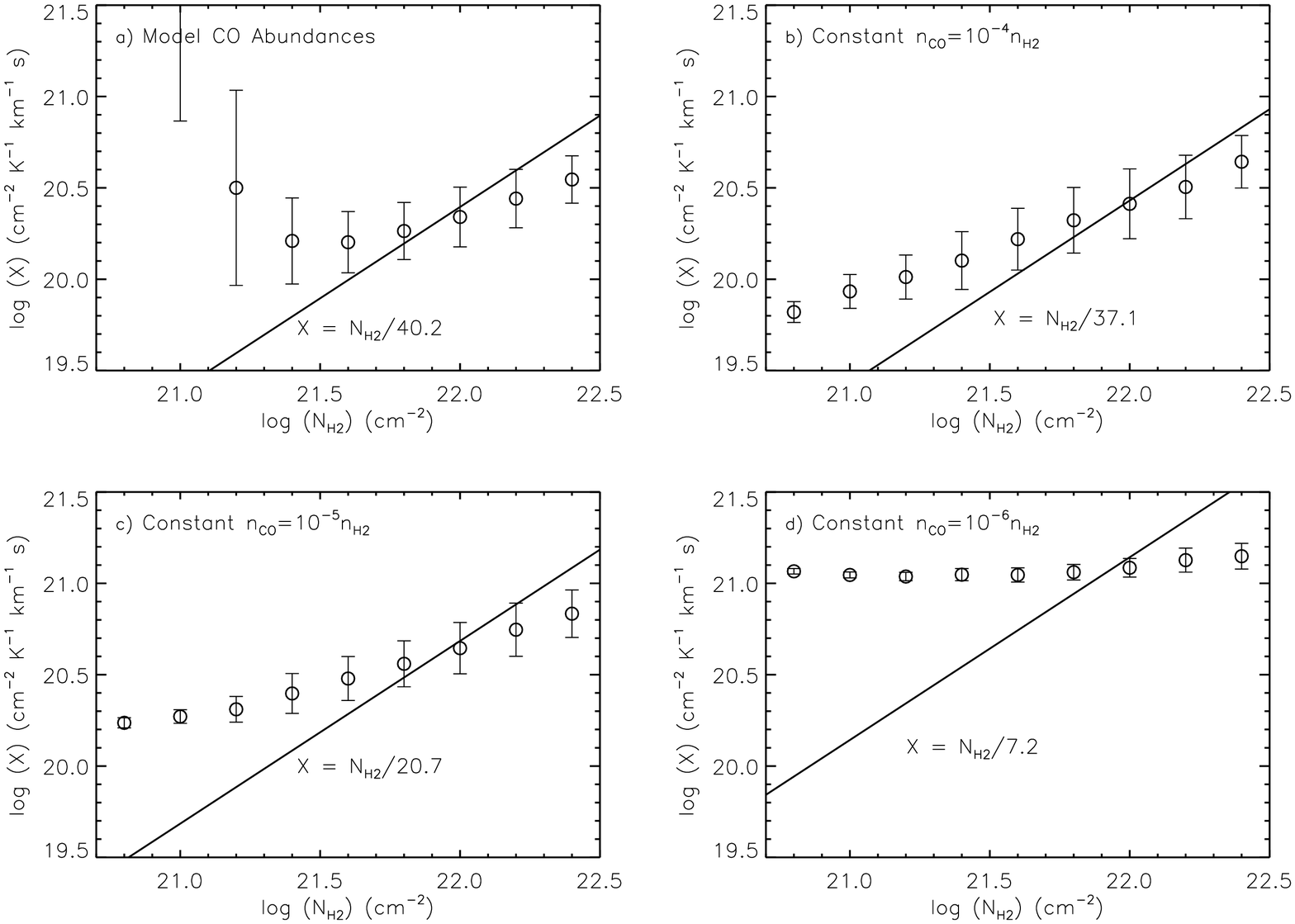} 
\caption{The \Xfac $-$\NHt relationship from model n300, as in Figure
  \ref{XvsN} (a), but with the CO number density reset to be (b)
  $10^{-4}$, (c) $10^{-5}$, and (d) $10^{-6} \times {\rm n_{H2}}$.
  Line in each plot shows $X=N_{H_2}/\overline{W}$ relation.  }
\label{constCOabund}
\end{figure*}

Figure \ref{constCOabund} shows the \Xfac\ - \NHt\ relation for (a)
the original model, as well as those with (b) constant \fCO
$=10^{-4}$, (c) $10^{-5}$, and (d) $10^{-6}$.  Also shown in each
panel is the line corresponding to the $X=N_{H_2}/\overline{W}$
relation, indicating complete saturation of the CO line.

With the original abundances, the $X=N_{H_2}/\overline{W}$
relationship is close to the model values only at the highest column
densities.  At densities \aplt 3$\times 10^{21}$ \cmsq, the
overplotted line underestimates \X.  In this regime, CO emission is
not saturated.  Figure \ref{constCOabund}b is analogous to the
limiting scenario where all the atomic carbon and oxygen is converted
to CO, so that \fCO =10$^{-4}$.  For this model, the
$X=N_{H_2}/\overline{W}$ relation is similar to the data above
\NHt\ $\sim 10^{21.5}$ cm$^{-2}$, indicating that the the CO line is
nearly fully saturated.  In Figure \ref{constCOabund}c-d, with lower
\fCO, line saturation becomes less and less important, especially at
lower column densities, finally resulting in a constant \X\ and
\NHt\ relation for \fCO = 10$^{-6}$.

In general, the CO line becomes saturated in regions with the highest
CO abundance.  With constant \fCO = $10^{-4}$, \X\ increases with
increasing \NHt\ everywhere, and the CO line is completely saturated
at column densities \apgt $10^{21.5}$ cm$^{-2}$
(Fig. \ref{constCOabund}b), resulting \X $\propto$ \NHt.  At lower CO
abundances, \W\ increases with increasing column density.  This
results in a shallower slope in the \X\ - \NHt\ relation.  At very low
\fCO = $10^{-6}$, $W$ is directly proportional to \NHt, so that \X\ is
constant at all \NHt.  Taken together, clouds with both low and high
CO abundances will tend to have a more limited range in \X\ than a
cloud with only large \fCO, which would have \X$\propto$ \NHt.

Notice that in Figure \ref{constCOabund}d, \X = $10^{21}$ \Xunits, for
the model with CO abundance \fCO=$10^{-6}$.  This is the quoted
abundance in diffuse Milky Way gas observed by \citet{Burghetal07} and
\citet{Lisztetal10}.  However, \citet{Lisztetal10} find \X $\approx
3\times10^{20} \approx$ \Xgal, which is a factor $\sim$5 lower than
the resulting value in Fig. \ref{constCOabund}d.  The discrepancy is
likely due to the combination of the differences in temperature and
linewidths between the ``diffuse'' MCs and more massive giant MCs. The
diffuse ISM has a higher temperature, up to $\sim$100 K.  As we
demonstrate in Section \ref{tempsec}, such high temperatures may
account for a significant fraction of the difference.  Further,
observations of low column density LoSs probably trace a larger volume
of the Galaxy, thereby including gas with a wider range in velocities
than those found in MCs.  As we discuss in the next section, larger
velocities may lead to lower \Xfac\ values.  Nevertheless, note that
the discrepancy may be partly due to the fact that in this numerical
experiment, we artificially fix the CO abundance.  The metallicity and
self-shielding are not self-consistently tracked in these experiments.

\subsection{X factor dependence on velocity}\label{velsec}

In order to test the sensitivity of the \Xfac\ to the velocity
\vlos\ or the integral over $dv$ appearing in Equation
\ref{Xfaceqnall}, we consider MC models with different velocity
fields.  In the MHD simulations, turbulence is generated by
continuously driving the gas velocities with uniform power between
wavenumbers $1 \le k \le 2$.  In the standard Milky Way MC model
presented so far, the saturation amplitude of the 1D velocity
dispersion is 2.4 \kms.

We test the effect of velocities by performing experiments on two
different sets of models for which the velocities differ from the
fiducial n300 model.  The first set of models have different driven
turbulent velocities.  In these experiments, the other parameters of
the simulations are identical to that from model n300 - only the
saturated-state (3D rms) velocities differ by factors of 0.2 and 2.
The second set of models are simply the original n300 models, but for
which the velocities in each zone are manually modified.  As discussed
in Paper I, the chemical evolution is not strongly affected by the
velocity field.  It is the metallicity, density, and background UV
radiation field which are the primary factors determining CO formation
\citep[see also][]{Gloveretal10, Glover&MacLow11}.  Given the
insensitivity of molecule formation to the velocity field, we can
directly modify the velocities in model n300.  We thus consider models
with chosen velocity dispersions, where all the other parameters are
equivalent to those in model n300.

\begin{figure}
\includegraphics[width=90mm]{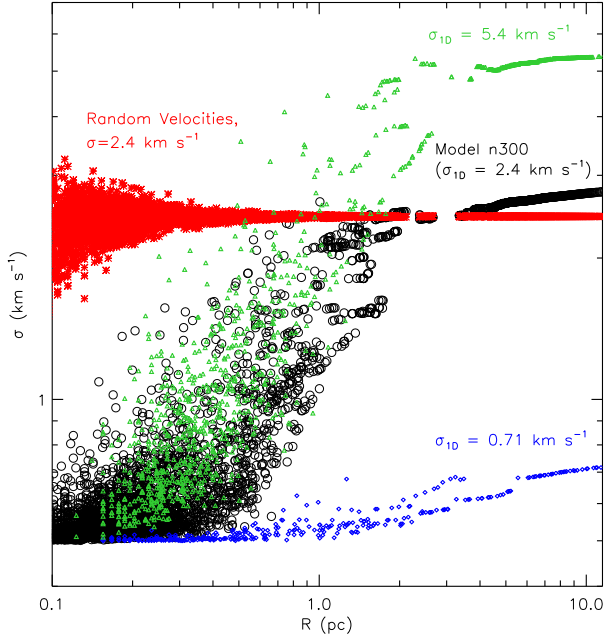} 
\caption
{The linewidth-size relationship from model n300.  Structures are
  identified directly from the 3D simulation.  The linewidths $\sigma$
  are computed by taking the dispersion of the (1D, or $v_z$)
  velocities from the identified structures in the simulations (black,
  blue, and green) or from a Gaussian distribution with a dispersion
  of 2.4 \kms\ (red).}
\label{lwsize}
\end{figure}

\subsubsection{Different levels of turbulence}
Figure \ref{lwsize} shows the linewidth-size relations for clumps in
the MHD models.  ``Clumps'' are identified directly from the
simulation (in the 3D density cube), through dendrograms
\citep{Rosolowskyetal08}.  The
dendrogram\footnote{https://people.ok.ubc.ca/erosolo/dendro.html}
algorithm identifies contiguous structures through iso-density
contours.  The linewidth of each clump is computed from the dispersion
in the corresponding region from the 3D ${v}_z$-cube.\footnote{We use
  the $\hat{z}$-component since the CO emission is calculated for the
  $\hat{z}$-direction.}  As evident in Figure \ref{lwsize}, the
linewidth-size relationship can be reasonably expressed as a power law
$\sigma \propto R^a$.  For the original n300 model (black circles) the
best fit exponent $a$=0.45.  This is in good agreement to the observed
linewidth size relationship \citep[e.g.][]{Larson81, Solomonetal87,
  Heyeretal09}.  At small radii approaching the resolution limit of
the simulation, the velocities approach a minimum threshold
corresponding to the microturbulent velocity of 0.5 \kms.  This chosen
value of the microturbulent velocity is greater than the thermal
velocity due to the observed linewidths at $\sim$0.1 pc scales.  At
large scales, the 1D linewidths $\sim$2.4 \kms\ are the overall
dispersion in the LoS velocities.

The blue and green points in Figure \ref{lwsize} show the $\sigma$ -
$R$ relationship from simulations with different forcing amplitudes,
with 3D \vrms $\approx$ 1 and 10 \kms, producing LoS velocity
dispersions of 0.71 and 5.8 \kms, respectively.\footnote{These LoS
  velocity dispersions include a contribution from the 0.5
  \kms\ microturbulent velocity, so are not exactly equal to
  $v_{\rm rms}/\sqrt{3}$.}  For these simulations, the best fit power laws
produce $a$ = 0.08 and 0.61, respectively.  The red points in Figure
\ref{lwsize} show the linewidth-size relationship from a model similar
to n300, but with the velocities drawn from a random distribution with
$\sigma$=2.4 \kms\ (which produces $a$=0).

\begin{figure*}
\includegraphics[width=120mm]{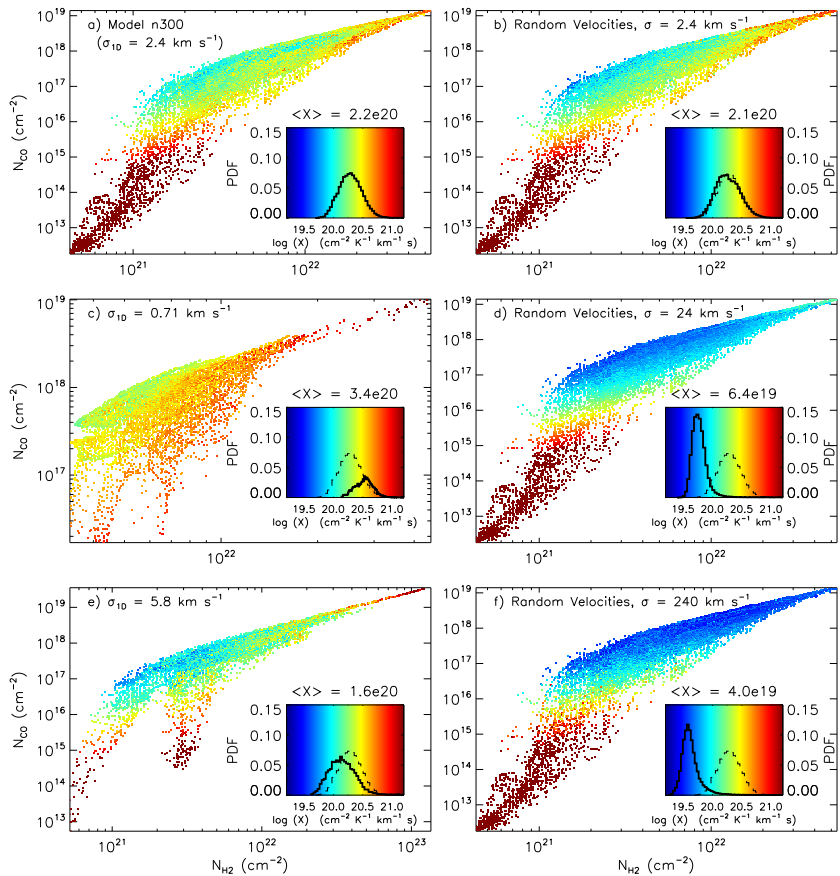} 
\caption
{The variation of the \Xfac\ with \NHt\ and \NCO, as in
  Fig. \ref{tvar}, but for models with different internal velocity
  fields.  a) Original model n300, which has $\sigma_{1D}$=2.4 \kms,
  b) model n300, but with velocities replaced from a Gaussian
  distribution with $\sigma$=2.4 \kms, c) model with $\sigma_{1D}$ =
  0.71 \kms\ (\vrms=1 \kms), d) n300, but with Gaussian velocities
  with $\sigma$=24 \kms, e) model with $\sigma_{1D}$ = 5.8
  \kms\ (\vrms=10 \kms), and f) n300, but with Gaussian velocities
  with $\sigma$=240 \kms.  In b)-e), the \Xfac\ distribution from the
  fiducial n300 model in a) is shown as the thin dashed histogram.
  The emission weighted \Xfac\ $\langle X\rangle $ is indicated in
  each panel.}
\label{Xvarvels}
\end{figure*}

Figure \ref{Xvarvels} shows the relationship between \X, \NHt, and
\NCO\ from the original n300 model, along with the models with
different velocity fields.  In Figure \ref{Xvarvels}b - f, the
velocities vary due to a different turbulent forcing amplitude, or
simply due to replacing the velocities from the n300 MHD simulation
with a Gaussian distribution.  As discussed, the original velocities
in model n300, (Fig. \ref{Xvarvels}a) have a LoS dispersion
of $\sigma$=2.4 \kms, due to turbulent forcing with 3D (rms) mean
dispersion \vrms = 5 \kms.

Figure \ref{Xvarvels}(c) and (e) show the \Xfac\ from models where the
turbulent forcing is varied from the original n300 to yield
$\sigma_{1D} \approx$ 0.71 \kms, or $\sigma_{1D} \approx$ 5.8 \kms,
respectively.  Figure \ref{lwsize} demonstrates that the velocities of
the large scale structures in these model vary by a factor of $\sim$3
above and below the original n300 model, but are rather similar on the
smallest scales due to the imposed microturbulence.  Because the
turbulent driving varies, the gas in these models contain different
amounts of CO, and different temperatures.  Comparing the other
properties of these models (in Table \ref{simprops}), the lower
velocities in the 0.71 \kms\ model result in lower temperatures
($\langle T\rangle _{vol} = 24$ K), and lower amounts of total CO (as
well as \Ht, $\langle n_{H_2}\rangle _{vol} = 127$ \cmt).  The 5.8
\kms\ model, on the other hand, has $\langle T\rangle _{mass} = 27$ K.
Nevertheless, the \Xfac\ for these models are both within $\sim 50\%$
of the n300 value ($\langle X\rangle=3.4\times 10^{20}$, $2.2\times
10^{20}$, and $1.6\times 10^{20}$ \Xunits, respectively, for
$\sigma_{1D} = $ 0.71 \kms, 2.4 \kms, and 5.8 \kms.).

\subsubsection{Purely Gaussian Velocities}
The right panels in Figure \ref{Xvarvels} shows the \Xfac\ from model
n300, but with the velocities replaced with random values drawn from a
Gaussian distribution with $\sigma$= b) 2.4, d) 24, and e) 240 \kms.
For the $\sigma$=2.4 \kms\ model, the linewidth-size relationship
differs significantly (Fig. \ref{lwsize}) from the fiducial model
n300.  The dispersions are similar on large scales, by design.
However, at smaller scales the dispersions in the original model
decreases, whereas there is a constant dispersion on all scales in the
modified model.  Nevertheless, the \Xfac\ distribution is very similar
to the original model n300.  The globally averaged \Xfac\ from the
original model is recovered.  That the \Xfac\ is very similar to the
original model suggests that {\it the details of the velocity
  structure, and its relationship to other physical properties such as
  mass or size, do not play an important role in determining the
  \Xfac.}  In particular, clouds with very different linewidth-size
relationships from $\sigma \propto R^{1/2}$ may produce \X
$\approx$\Xgal.

In models with random velocities having 1D dispersions $\sigma$=24 and
240 \kms, the \Xfac\ is systematically lowered, to $\langle
X\rangle=6.4\times 10^{19}$ and $4\times 10^{19}$ \Xunits,
respectively.  {\it Therefore, the integrated CO intensity $W$ is not
  sensitive to the velocity structure, but only the extent of the
  range in velocities}.  This occurs because with larger velocity
differences between regions along a LoS, more CO line photons are able
to escape the cloud and ultimately be detected, resulting in an
increase in \W.  Increasing \deltav\ thereby effectively reduces the
percentage of mass that has $\tau > 1$ (and is therefore
``invisible''), so that more CO becomes visible.  Notice that the
difference in the \Xfac s of the models with dispersions of 24 and 240
\kms\ (Fig.\ref{Xvarvels}d and f) is modest.  Thus, \X\ does not
simply scale inversely with $\sigma$.  In fact, the results from
models with $\sigma=2.4, 5.8$, and 24 \kms show a behavior closer to
$X \propto \sigma^{-1/2}$ than $X \propto \sigma^{-1}$.  We return to
this point in Section \ref{cldavespec}.

Of course, the range in velocities must be sufficiently well-sampled,
so that there are no significant velocity gaps, which will indeed be
the case in MCs.  The $\sigma$=24 and 240 \kms\ models have a large
range in velocities, and consequently, likely have under-resolved
velocity gradients.  In the Appendix, we discuss how radiative
transfer calculations may provide inaccurate intensities in regions
where the velocity gradients are poorly resolved, and how our analysis
accounts for this effect.

\subsection{Does ``cloud counting'' result in a constant X factor?}\label{specsec}

One explanation for the lack of variation in the \Xfac\ in the Galaxy
is that the integrated CO intensity is a measure of the number of
optically thick ``cloudlets'' along the LoS, hereafter LoS.
This is known as the ``mist'' model proposed by \citet{Solomonetal87}.
Since we have knowledge of all the relevant quantities affecting the
\Xfac, we can test this idea by inspecting the characteristics of the
gas contributing to the observed emission.

\begin{figure*}
\includegraphics[width=120mm]{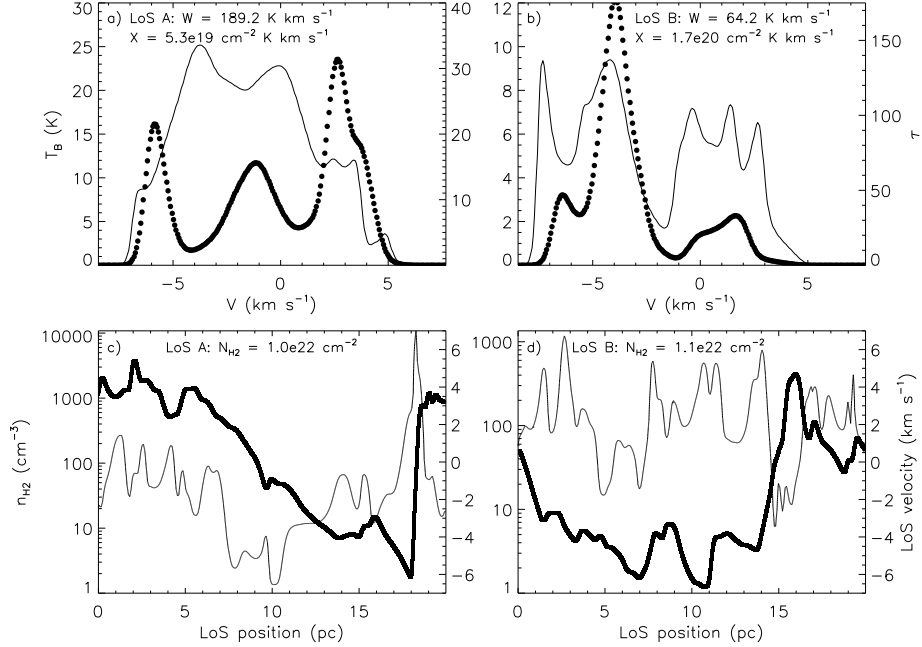} 
\caption
{Top panels - CO spectra (lines - left axis)) and optical depth
  (circles - right axis), and bottom panels - \Ht\ column densities
  (thin - left axis) and LoS velocities (thick - right axis) from two
  LoSs in the n300 simulation.  The corresponding integrated
  intensities and \Xfac s are listed in a-b, and the column densities
  are listed in c-d.}
\label{spec1}
\end{figure*}

Figure \ref{spec1} shows the spectrum from two LoSs.  The top panels
show \Tb\ and $\tau$ at each spectral channel.  The bottom panels show
the corresponding volume density and velocity.  For reference, the
observer is situated beyond LoS position 0 in panels c-d, so that
observed line photons are traveling from high LoS position towards
lower LoS position (to the left on the plots).

Both LoSs A and B have similar total column densities, 1.0 and 1.1
$\times10^{20}$ \cmsq, respectively.  However, the integrated
intensities differ by a factor of $\sim 2.5$.  Perhaps surprisingly,
LoS A, which has slightly lower column density, has a much higher
intensity.  As a result of this difference in \W, the resulting \Xfac
s for these LoSs as computed from Equation \ref{Xfaceqn} also differ.

The differences in the total intensities and line shapes from the LoSs
depicted in Figures \ref{spec1}a-b can be understood by inspecting the
density and LoS velocity distribution shown in Figures \ref{spec1}c-d.
LoS A has relatively low density gas \aplt 100 \cmt\ in the majority
of positions along the LoS, at position 0-17 pc.  Near LoS position 17
pc, there is a sharp jump in density, and an associated perturbation
in velocity.  The lowest LoS velocities associated with this shock,
\aplt $-$5 \kms, are unique to this particular region along the LoS -
no other region along this LoS has similar velocities.  Accordingly,
observed emission in the $-$5 to $-$7 \kms\ range originates from this
high density, optically thick cloudlet.  Emission at velocities $-$3
to 4 \kms\ from this cloudlet may be absorbed by the gas lying along
the LoS with similar velocities.  However, this gas has very low
column density, and so in fact does not significantly attenuate the
emission from the shock.  Thus, the whole line profile in the $-$7 to
3.5 \kms\ range is due to the high density shock.  Only the weak
remaining emission at velocities \apgt 4 \kms\ is due to gas at
positions \aplt 3 pc.  We discuss some numerical effects in regions of
such high velocity contrast in the Appendix.

The LoS in Figures \ref{spec1}b does not have any gas with \nHt $>$
2000 \cmt.  There are numerous optically thick regions, with a
significant overlap in velocities.  Some regions, such as the peak
near position 8 pc, have small velocity gradients, while others
contain a large range in velocities, as the peak near position 14 pc.
Due to the large overlap in velocities, especially at $-$4 \kms, the
integrated optical depth reaches very high values up to 170.  The
combination of velocity overlap and lower density gas results in LoS B
having a lower integrated intensity than LoS A.

\begin{figure*}
\includegraphics[width=120mm]{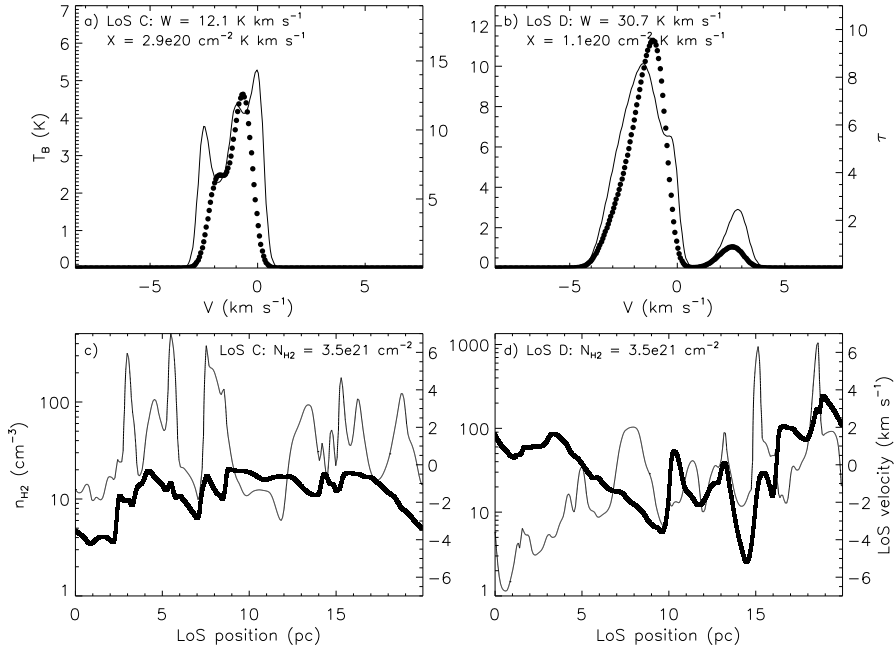} 
\caption
{Similar to Fig. \ref{spec1}, but for different LoSs.  The column
  density of both LoSs are equivalent.}
\label{spec2}
\end{figure*}

Figure \ref{spec2} shows two other LoSs with lower column densities;
these have equivalent \NHt\ = 3.5$\times 10^{21}$ \cmsq.  Yet, the
integrated intensity varies by a factor of $\sim$3, resulting in an
equivalent discrepancy in the \Xfac.  Judging by the optical depth
profile, there appears to be either one or two cloudlets.  However,
the detailed velocity and density profiles show that the structure is
much more complex.

Almost all of the gas along LoS C has densities \apgt 10 \cmt.  Most
of the gas velocities, especially those associated with the density
peaks with \nHt \apgt 100 \cmt, lie in the range $-$3 to $-$1 \kms.
There are many density peaks with overlapping velocities, but the
resulting line profile has only three peaks.  Since $\tau >$ 1, the
observed intensity at a given velocity emerges from the last density
peak along the LoS with the given velocity.  Thus, much of the
emission from LoS position $>$ 10 pc is absorbed.  By rerunning the
radiative transfer on this LoS by excluding some of the gas, we have
verified that a significant number of CO line photons are absorbed.
We provide examples demonstrating this scenario in Section
\ref{cldavespec}.

For LoS D, a fraction of the gas has \nHt \aplt 10 \cmt.  This LoS has
two well defined peaks in the line profile.  These peaks are centered
on $-$3 and 3 \kms.  At those velocities, there are two distinct
cloudlets with \nHt\ \apgt 1000 \cmt.  Clearly, the emission from LoS
D comes from the most dense regions along the LoS.  Since they are
well separated in velocity, both are easily detected and contribute to
a larger integrated intensity than LoS C.

The comparison of observed profiles, densities, and velocities in
Figures \ref{spec1} and \ref{spec2} suggest that the simple ``mist''
model does not accurately capture the complexity intrinsic to line
radiative transfer from a turbulent medium.  In the comparison of LoS
A and B in Figure \ref{spec1}, both of which have a large \NHt, A has
one true cloudlet with very high density.  This cloudlet is only
present in a localized region along the LoS, but contains a large
range in velocities, and therefore is the source of most of the
observed emission.  Other emitting regions with different velocities
only contribute slightly to the spectrum, due to their significantly
lower densities.  LoS B has numerous lower density cloudlets, many of
which have similar velocities.  The integrated intensity is lower,
though the total amount of gas along this LoS is slightly larger than
LoS A.

In the comparison of low density LoSs in Figure \ref{spec2}, both LoSs
have a few regions with clear density peaks.  However, the dominant
cloudlets in LoS D have larger density, and span a larger range of
velocities.  Thus, LoS D has a higher total intensity than LoS C even
though the total column density in both LoSs are equivalent.

The ``mist'' model would predict that LoSs with more high density
cloudlets would have higher integrated intensities, since the
cloudlets are separated in velocity.  However, comparison of LoS A and
B show the opposite: one very dense cloudlet with a large velocity
gradient may be responsible for the whole LoS profile.  This
integrated intensity \W\ may be larger than that from a different LoS
with a larger number of cloudlets, but nevertheless a similar total
column density and total velocity width.  Further, two LoSs with
equivalent \NHt\ and similar number of cloudlets may have different
\W\ if one LoS has overlapping cloudlets in velocity space, while the
other LoS has well distributed cloudlets in velocity space.  This
effect is partly at work in the comparison of LoS C and D in Figure
\ref{spec2}.

We have shown that in a turbulent medium, emitting cloudlets do in
fact overlap in velocity space \citep[see
  also][]{Ballesteros-Paredes&MacLow02, Shettyetal10}, and that
individual cloudlets may have a range in velocities, thereby
dominating the emission at all velocities in that LoS.  These factors
all compromise any simple relation between \W\ and \NHt, and therefore
any direct scaling between \X\ and \NHt.  Of course, this analysis
considers individual LoSs through a molecular cloud, whereas the idea
of a constant \Xfac\ is usually discussed in the context of whole
molecular clouds.  Accordingly, we next consider the cloud-averaged CO
spectrum.

\subsection{Cloud-averaged \Xfac\ and spectra}\label{cldavespec}

\begin{figure}
\includegraphics[width=90mm]{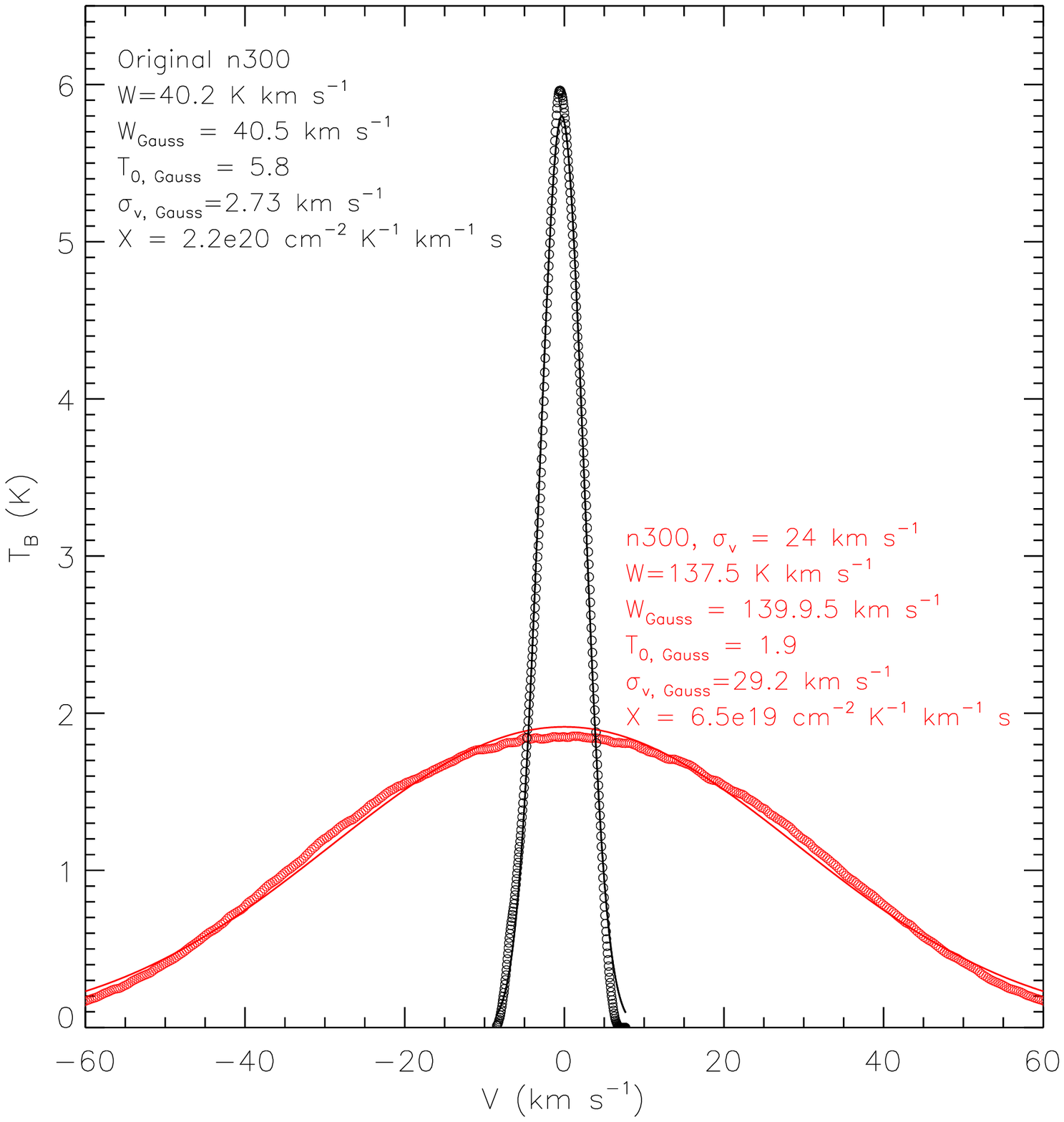} 
\caption
{Averaged spectrum from model n300 (black), and model n300 for which
  the velocities were replaced from a random distribution with
  $\sigma$ = 24 \kms (red).  Lines show best fit Gaussians.
  Integrated intensity, fit Gaussian parameters, and \Xfac, are listed
  for both models.}
\label{gausfits}
\end{figure}

Figure \ref{gausfits} shows the averaged spectrum for the whole n300
cloud model, as well as model n300 where the velocities are are
manually replaced with $\sigma$ = 24 \kms\ (see
Fig. \ref{Xvarvels}d).  Each point shows the mean brightness
temperature of each channel in the synthetic observation.  The
integrated intensity of this spectrum is \W = 40.2 and 137.5
\Wunits\ for the fiducial and velocity-altered model, respectively.

The best fit Gaussians to the averaged spectra are also shown in
Figure \ref{gausfits}.  The spectra are very well fit by Gaussians.
Using the peak and dispersion of the Gaussian fit, the corresponding
integrated intensities reproduce the true integrated intensities.
Using \W\ along with the mean column density $\overline N_{H_2} =9.0
\times 10^{21}$ \cmsq\ in Equation \ref{Xfaceqn} results in \X = 2.2
$\times10^{20}$ and 6.5$\times10^{19}$ \Xunits\ for the original and
24 \kms\ models, respectively. These values of the \Xfac\ are
equivalent to the emission-weighted mean \X\ found in Section 4.1 and
listed in Figure \ref{Xvarvels}.  The \Xfac\ value for n300 is nearly
identical to the reference value in Table \ref{simprops}.  For the
altered velocity model, $\langle v_z^2\rangle _{mass}$ = 24 \kms\ so
the reference $\langle X\rangle _{ref} = \langle n_{H_2}\rangle _{vol}
L / [\langle T\rangle _{mass} (\langle v_z^2\rangle
  _{mass})^\frac{1}{2}]$ = 1.9 $\times 10^{19}$ \Xunits.  This value
is significantly lower than \X\ determined by the spectrum, largely
due to the fact that $T_B/\langle T\rangle_{mass}$ is a factor 3 lower
for the 24 \kms\ model than for model n300 with $\sigma_{1D}$=2.4
\kms.  {\it Therefore the mass-weighted quantities $\langle T\rangle
  _{mass}$ and $\langle v_z^2\rangle _{mass}^{1/2}$, cannot fully
  account for \W.}  We note that the line widths of the observed
spectrum are slightly larger than the intrinsic velocity dispersion.
This is due to some extent to our choice of the microturbulence, which
we discuss in more detail in the Appendix.

The cloud-averaged spectra in Figure \ref{gausfits} show very smooth
profiles, and do not demonstrate any of the variability seen along
individual LoSs.  Clearly, the cloud with a larger velocity dispersion
is brighter because the emission from more gas with a larger
dispersion is able to escape the cloud.  Even though both models have
the same mass, their integrated emission differs by a factor 3.4.
This is due to a combination of a larger velocity dispersion (by a
factor $\sim 10$), and a lower peak brightness temperature (by a
factor $\sim 3$) for the 24 \kms\ model compared to model n300.  As a
result, use of the integrated intensity alone, or any \Xfac, would
provide inaccurate mass estimates.

We have demonstrated that there is no simple relationship between
\W\ and \NHt\ in highly turbulent clouds (see also Paper I).  This
lack of correlation is due to the combination of the optically thick
nature of CO, and the complexity in the structure of CO bright
regions: emitting gas along individual LoSs sometimes overlaps in
velocity space and sometimes does not, and even lower density gas not
necessarily found in cloudlets contributes significantly to the
observed intensity.  Under such circumstances, there should be
significant amounts of gas that is untraceable in the CO line.

\begin{figure}
\includegraphics[width=90mm]{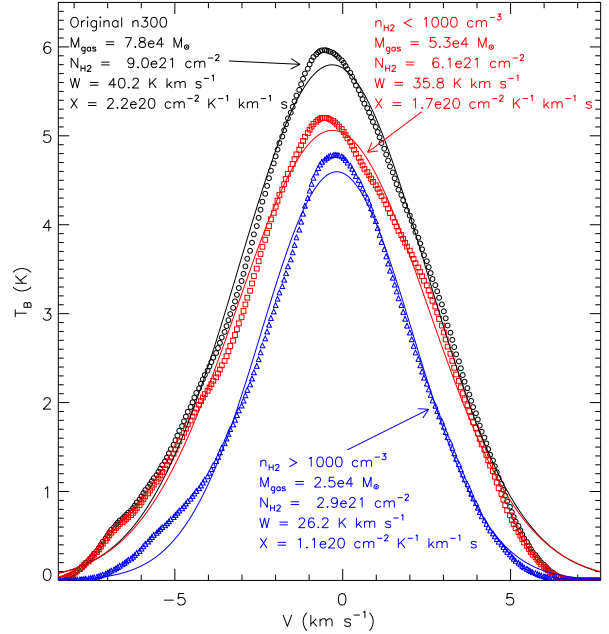} 
\caption
{Averaged spectrum from original model n300 (black circles), along
  with spectrum from the original model but only gas with \nHt $<$
  1000 \cmt (red squares) and $>$ 1000 \cmt (blue triangles) is
  considered.  Lines show best fit Gaussians.}
\label{carve}
\end{figure}

Figure \ref{carve} shows the averaged spectrum of three models.  The
black spectrum is from the original n300 model.  The red and blue
points show the results of the radiative transfer calculations from
model n300, but only when gas with \nHt $<$ 1000 \cmt\ and $>$ 1000
\cmt\ is included in the calculation, respectively.  The CO and \Ht\
density is set to zero in zones with \nHt $<$ 1000 \cmt or $>$ 1000
\cmt, respectively.

The gas mass in the model that has \nHt $<$ 1000 \cmt\ is 5.3$\times
10^4$ \msun, which is roughly 2/3rds of the total mass in the original
n300 model.  However, the integrated intensity from this model is 90\%
of that from model n300.  The total mass at \nHt $>$ 1000 \cmt,
2.5$\times 10^4$ \msun, accounts for the remaining 1/3 of the mass in
model n300.  But the integrated intensity from this model is 26.2
\Wunits, which is 65\% of the total \W\ of the original model.

Due to the significant amount of self-absorption occurring in both low
density and high density gas, removing some gas simply allows other
gas along the LoS to become ``visible.''  There are LoSs
where some CO emitting regions go completely undetected, since other
optically thick gas lying in the same LoS with similar velocities will
absorb its emission.  Consequently, a spectrum from a given LoS may
depend on the observing direction.  For example, if the observer were
located at LoS position $>$20 pc in Figures \ref{spec1} and
\ref{spec2}, instead of at LoS position $<$0 pc, the spectra may have
different profile shapes.  We have indeed found such discrepancies in
many LoSs.  This is further confirmation that there is no simple
correlation between \NHt and \W.  We stress again that the use of a
constant \Xfac\ for numerous clouds will not provide reliable
estimates of the molecular mass.

\section{Discussion}\label{discsec}

\subsection{Why is the Galactic \Xfac\ found to be nearly constant?}

After investigating the dependence of the \Xfac\ on the physical
characteristics of synthetic MCs, we are now in a position to
interpret the observed trends, as well as compare our results with
previous efforts.

For our model with Milky Way parameters, we have demonstrated that
under certain conditions, using the mass-weighted mean values of
temperature and velocity dispersion may provide reasonably good
estimates of \Xfac.  Thus, it might seem unsurprising that systems
with similar mass-weighted temperatures or velocity dispersions are
able to reproduce the \X\ factor from the original model.  By
exploring models with different temperatures and velocities, however,
we find that $X_{ref} = \langle N_{\rm H_2}\rangle / (\langle
T\rangle_{mass} \, \sigma_{v,los}) $ does not in general agree with
the value of $\langle X\rangle$ =\NHt/\W\ computed from radiative
transfer models.  Further, $\langle X\rangle$ does not simply depend
on the inverse of the temperature or velocity.

Even though the experiments testing the variation of a constant
kinetic temperature on the CO line do not self-consistently follow the
coupling of other MC properties with $T$, we may compare our results
with previous investigations.  We find that \X $\, \propto T^{-0.5}$,
which is a weaker \X $- T$ dependence than the \X $\, \propto
T^{-1.3}$ behavior found by \citet{Kutner&Leung85}.  Their models have
velocities which are dominated by microturbulence, but nevertheless
obey the empirical linewidth-size relationship.  We find that
microturbulence does not significantly affect the integrated
intensity, nor correspondingly the \Xfac\ (see Appendix showing
profiles with different microturbulent velocities).  In our models,
the CO intensity is more sensitive to the macroturbulent velocities.

Thus, our results are in closer agreement to the investigation by
\citet{Wolfireetal93}, who find that macroturbulent or clumpy models
are better able to reproduce observed line profiles than
microturbulent models.  However, a key difference between our model
and that of \citet[][as well as
  \citealt{Dickmanetal86}]{Wolfireetal93} is that in their clumpy
models, much of the CO emission emerges from distinct optically thick
clumps.  Further, their models require approximately one clump along
each LoS at a given velocity in order to reproduce observed spectra.

The \citet{Wolfireetal93} models are conceptually similar to the
``mist'' model often employed to explain the constant
\Xfac\ \citep{Dickmanetal86, Solomonetal87} and described in Section
4.5.  In our turbulent models, only the cloud-averaged CO linewidths
provide accurate measures of the total (1D) velocity dispersion.
Along an individual LoS, the line shape is strongly dependent on the
density and velocity structure.  We find that though dense gas is
responsible for much of the observed CO emission globally, along
individual LoSs diffuse gas may also be a significant source.
Depending on the gas velocities along the LoS, lower density regions
may also be optically thick, and so may contribute substantially to
the emergent intensity.

One of the key results of our work is that the \Xfac\ is insensitive
to the detailed linewidth $-$ size relationship in the cloud.  In
Section 4.4 we demonstrated that a cloud consisting of gas with random
velocities has an almost identical \Xfac\ distribution with a cloud
that has a well defined $\sigma \propto R^\frac{1}{2}$ linewidth-size
relationship.  This suggests that the linewidth-size relationship has
very little influence in determining the \Xfac, or \W.  The lack of
dependence of \W\ on the detailed velocity structure suggests the use
of a representative velocity width for the $\int dv$ term in Equation
\ref{Xfaceqnall}.  Although linewidths computed from the mean spectrum
may differ slightly from $\langle v_z^2\rangle _{mass}^{1/2}$, a more
significant effect of radiative transfer is to reduce $T_B$ when the
velocity width increases.  Consequently, we find that \W\ does not
increase linearly with the velocity dispersion.  In particular, we
find that an increase of the velocity width by a factor 10 or 100 only
reduces $\langle X\rangle$ by a factor 3 or 5 relative to model n300
(see caveats in Appendix).

Along with \NHt, \W\ determines the \Xfac\ through its traditional
definition given by Equation \ref{Xfaceqn}.  Based on our tests, for
\NHt\ $\sim 10^{22}$ \cmsq, the \Xfac\ can only vary by a factor \aplt
5, given the limited range of dense-gas temperatures 10 - 25 K and
velocities $\sim$2 - 5 \kms\ observed for Galactic MCs
\citep[e.g.][]{Solomonetal87, Roman-Duvaletal10}. The mass-size
relationship proposed by \citet{Larson81} implies a constant mean
column density in MCs.  Recent observations have challenged the notion
that all MCs have column densities, averaged over large scales,
similar to one another \citep{Heyeretal09,Kauffmannetal10b}.
Nevertheless, column densities are not expected to vary significantly
from cloud to cloud, and they generally fall in the range from
$10^{21} - 10^{22}$ \cmsq.  Structures with much lower column
densities than $\sim 10^{21}$ \cmsq\ probably would not be observed as
{\it molecular} clouds, as there would be insufficient CO to be
detected.  Clouds with column densities much larger that $\sim10^{22}$
\cmsq\ would be very gravitationally unstable - rapid collapse,
fragmentation, and the destructive effects of star formation would all
prevent clouds from maintaining such high column densities over long
periods of time (within an environment of much lower mean density and
column density).  Thus, we believe the limited range in \NHt, velocity
dispersion, and MC temperature together explain the limited range in
the \Xfac\ found by numerous observations of Galactic MCs.

\subsection{Does a constant \Xfac\ require virialized clouds?}

A common interpretation of the limited range in \Xfac s from
observations of Galactic MCs is that they are ``virialized'' objects.
The ``mist'' model of \citet{Solomonetal87} invokes the ``virialized''
cloud depiction to explain the limited range in the measured virial
mass - CO luminosity (\MtoL) ratio.  Generally, a mass-to-luminosity
ratio, $\alpha$, is simply the \Xfac\ written in terms of a gas mass
and CO luminosity, instead of \NHt\ and \W, so that \X $=4.5 \times
10^{19} \alpha$ if $\alpha$ has the units \msun /(\Wunits pc$^{-2}$).
Since the derivation of a constant $\alpha_{VT} \equiv$\MtoL\ for
virialized clouds has been presented in numerous works
\citep[e.g][]{Solomonetal87, Maloney90, Young&Scoville91}, we only
briefly outline these arguments, and then discuss how our models
relate to such a description.

For a cloud in virial equilibrium, the internal velocities are fully
governed by the gravitational field due to the gaseous mass,
\begin{equation}
\sigma_{VT}^2 \equiv \frac{G M_{VT}}{5R}.
\label{virvels}
\end{equation}
Here, we have used ``VT'' to label the one-dimensional virial velocity
width $\sigma_{VT}$, and the factor 5 assumes a spherical, uniform
cloud \citep{Bertoldi&McKee92}.  Note that in this notation,
$\sigma_{VT}$ is the 1D velocity dispersion (assuming an isotropic
distribution), whereas $\sigma$ refers to the observed linewidth along
the LoS.

Cloud-averaged spectra often have Gaussian shapes, so that the CO
luminosity can be computed using the Gaussian line parameters
$T_{B,0}$, the peak brightness temperature, and linewidth $\sigma$,
along with the projected area $\pi R^2$:
\begin{equation}
L_{CO} =\sqrt{2\pi}T_{B,0} \sigma \pi R^2.
\label{gauslum}
\end{equation}
Taking Equations \ref{virvels} and \ref{gauslum} together, and assuming
$\sigma=\sigma_{VT}$,
\begin{equation}
\alpha_{VT} \equiv \frac{M_{VT}}{L_{CO}} \propto \frac{\sigma/R}{ T_{B,0}}.
\label{ml}
\end{equation}
Using Equation \ref{virvels}, $\sigma/R \propto (GM/R^3)^{1/2}$ so
that Equation \ref{ml} is equivalent to
\begin{equation}
\alpha_{VT} \equiv \frac{M_{VT}}{L_{CO}} \propto \frac{\rho^{1/2}}{T_{B,0}}.
\label{lumlrho}
\end{equation}
If $\rho$ and $T_{B,0}$ take on a limited range of values, this yields
an approximately constant $\alpha_{VT}$ and \X. 

When further combined with an empirical linewidth-size relation
$\sigma \propto R^{1/2}$, equation \ref{virvels} yields
\begin{equation}
M_{VT} \propto \sigma^4,
\label{mvels}
\end{equation}
and equation \ref{gauslum} yields $L_{CO} \propto T_{B,0}
\sigma^5$.  These relations combine to give 
\begin{equation}
M_{VT} \propto \left(\frac{L_{CO}}{T_{B,0}}\right)^{0.8}.
\label{MasL}
\end{equation}

CO observations are often used to obtain cloud masses through Equation
\ref{virvels}, assuming $\sigma = \sigma_{VT}$.  Such an analysis usually
produces a strong correlation between $M_{VT}$ and $L_{CO}$, with a
power law index 0.8.  Though we have only followed proportionalities
in Equations \ref{ml} - \ref{MasL}, it is straightforward to consider
the coefficients explicitly
\citep[e.g.][]{Maloney90,SolomonVandenBout05}.  The observed $M_{VT}
- L_{CO}$ trend also results in good agreement in these coefficients.
The agreements in both the power law exponents and coefficients are
often taken as evidence that MCs are virialized objects, and that
\NHt\ is approximately constant for MCs, since $N_{H_2}\propto \Sigma
\propto M/R^2 \sim$ constant if $\sigma^2 \propto M/R$ and $\sigma
\propto R^{1/2}$.

The apparent agreement between the observed \MtoL\ and the analytical
derivation outlined above dependends on two crucial assumptions, the
mass-size and linewidth-size relations.  As discussed above, numerous
observations demonstrate a $\sigma \propto R^{1/2}$ relationship over
a range of scales for MCs \citep{Larson81, Solomonetal87,
  Bolattoetal08}.  However, recent analysis by \citet{Heyeretal09} of
higher resolution observations of the clouds studied by
\citet{Solomonetal87} found that the coefficient of the linewidth-size
relationship may not be constant.  Furthermore, the observed mass-size
relationship $M \propto R^{2}$ is rather uncertain.  Using
observations of MCs in $^{12}$CO and $^{13}$CO, which should trace
more of the cloud gas due to the lower optical depth,
\citet{Roman-Duvaletal10} find an exponent of 2.4 rather than 2.  By
assessing the mass on a continuous range of scales in a number of
Galactic MCs, \citet{Kauffmannetal10a, Kauffmannetal10b} found no
universal power law mass-size relationship.  Additionally, when
optically thin (e.g. dust) observations are employed, projection
effects may lead to overestimated masses
\citep{Ballesteros-Paredes&MacLow02, Gammieetal03, Shettyetal10}.
Another potential issue is that the definition of cloud size $R$ may
affect any $M$ estimates.  Usually, the size $R$ is set to the radius
of a circle with area equivalent to that of the projected area of the
cloud.  Using a different definition of $R$ may result in an altered
mass-size relation, and possibly even a different linewidth-size
relation.  These caveats will all affect any derived mass - luminosity
correlations.

In this paper, we have studied CO emission from turbulent MC models
which do not have self-gravity, and thus are not ``virialized.''
Nevertheless, the virial mass inferred from the synthetic CO
observation accurately recovers the cloud mass (7.8$\times10^{4}$
\msun) to within 1\%, for the n300 model.  Since the parameters
adopted for the n300 model were designed to represent Milky Way MCs,
this agreement may lead to the spurious inference that the MC is
``virialized.''  Additionally, we have found that the integrated
emission, and hence the \Xfac\ for the adopted value of \NHt, is
insensitive to the velocity structure (see Section 4.4).  As a result,
a cloud with $\sigma \gg \sigma_{VT}$ or $\sigma \ll \sigma_{VT}$
which is far from ``virialized'' could still have quite similar
\X\ values to a cloud with $\sigma \sim \sigma_{VT}$.  Consequently,
our analysis supports the arguments put forth by \citet{Maloney90} and
\citet{Combes91} that a constant \Xfac\ does not require MCs to be
``virialized.''

In effect, our analysis shows that \X$\sim$\Xgal\ is simply a result
of the limited range of both the numerator and denominator of Equation
\ref{Xfaceqnall}, or of the parameters entering the right hand side of
Equation \ref{ml} or \ref{lumlrho} for gravitationally-bound MCs: the
densities, brightness temperatures, and (observed) velocities.  Only
{\it if the velocities are governed solely by gas self-gravity through
  Equation \ref{virvels} and a universal linewidth-size $\sigma
  \propto R^{1/2}$ relationship holds can a (nearly) constant
  \MtoL\ ratio be expected for all MCs.}  The linewidths of observed
clouds, which measure the internal motions, may or may not be set
solely by gas self-gravity.  An additional caveat is that the Equation
\ref{virvels} may not be sufficient for indentifying the dynamical
state of a cloud, since it does not account for surface, magnetic and
time-varying terms in the virial equation
\citep{Ballesteros-Paredes06}.

\subsection{Future work: \Xfac\ in molecular dominated regions}

For molecular environments different from Milky Way GMCs, the
different (volume and column) densities, velocities, and/or
temperatures, should contribute to a variation in the corresponding
\Xfac s.  As indicated in the previous section, we have focused on one
model with densities similar to those of Milky Way MCs.  Molecular
dominated regions such as ULIRGs and the Galactic center have much
higher column densities, in addition to elevated temperatures and
observed velocity dispersions, so the results found in this work are
not directly applicable to those sources.  In this section, we briefly
explore the results from the previous section in the context of
ULIRGs, and motivate an analogous study of higher column density
models.

In molecular dominated regions, the \Xfac\ is believed to be lower
than \Xgal\ (see e.g. Fig. \ref{Xenvirons}, \citealt{Tacconietal08}).
In part, this is because the standard value of \Xgal\ yields molecular
masses that are sometimes larger than the dynamical masses of the
galaxies.  In order to avoid this unphysical situation,
\citet{Downesetal93} proposed that the masses traced by CO
observations of ULIRGs are not the virial masses, but rather the
geometric mean of the gas and dynamical masses, i.e. $L_{CO} \propto
T_{B,0} (M_{gas} M_{dyn}/\rho_{gas})^{1/2}$.  In their interpretation,
the CO linewidths trace the dynamical masses of the starbursting,
molecular rich systems.  Thus, the virial velocities and masses in
Equation \ref{virvels} must be replaced by the dynamical velocities
and masses of ULIRGs.  Instead of using $\alpha_{VT} L_{CO}$ (Eqn.
\ref{lumlrho}) to estimate masses, one can estimate the product of
dynamical and gas mass of ULIRGs through $(\alpha L_{CO})^2$, where
$\alpha = 2.6 n_{gas}^{1/2} /T_{B,0}$ \citep[Eqn. 5
  in][]{SolomonVandenBout05}.  The ratio $M_{gas}/L_{CO}$ computed
from ULIRGs is $\sim 3-6$ times lower than that estimated from MCs
\citep[e.g.][]{SolomonVandenBout05}.

ULIRGs have observed CO linewidths of the order of a few hundred
\kms\ \citep[e.g.][]{Solomonetal97,Downes&Solomon98}.  Since the
molecular component of ULIRGs are inclined rotating disks, the
observed CO linewidths are primarily tracing the rotational motion of
the molecular-dominated ISM.  \citet{Downes&Solomon98} find that
turbulent velocities of ULIRGs must be significantly higher than
Galactic MCs in order to accurately fit their observed lines.
However, due to the compact extent of ULIRGs and the high optical
depth of CO, accurately separating the turbulent and rotational
contributions to the observed linewidths is not trivial.

The observed brightness temperatures of ULIRGs are significantly lower
than Galactic MC values.  However, ULIRG kinetic temperatures can be
\apgt 10 times larger due to enhanced star formation activity
\citep[e.g.][]{Solomonetal97}.  In Section \ref{cldavespec}, we showed
that the brightness temperature indeed decreases with increasing
velocity dispersion.  Accurately quantifying this \Tb $-\sigma$
relationship calls for more realistic models which self-consistently
contain higher temperature, higher dispersion, and have sufficiently
high resolution (see Appendix).

In Section \ref{velsec}, we showed that a model with random velocities
with dispersion 240 \kms, similar to the values found from ULIRGS,
produces \X = 4.0$\times 10^{19}$ \Xunits.  In addition, we found \X =
7.9$\times 10^{19}$ \Xunits\ for the T=100 K model from section 4.2.
These \Xfac\ values are a factor $\sim$3-5 lower than \Xgal\ and the
value found from the original n300 model, and are consistent with
\Xfac\ estimates from ULIRG observations.  These could be due to
either extremely high small-scale turbulent velocity dispersions (as
experienced by photons escaping from the molecular disk) and moderate
gas temperatures, or much higher gas temperatures and moderate
small-scale turbulence combined with very large non-circular motions
on $10^2-10^3$ pc scales (still within the typical beam size for
molecular observations, but larger than the molecular disk thickness).
We note that none of our current models has column densities
comparable to ULIRGS.  The effects of higher star formation, very high
density and total column density, and rotation all need to be studied
in detail to understand the CO-\Ht\ conversion factor from ULIRG
systems.  More detailed modeling should quantify the contribution of
rotational and turbulent velocities, along with other physical
properties in setting the \Xfac\ of ULIRGs.  Efforts considering such
environments are currently underway (current authors, and Narayanan et
al. in prep).

\section{Summary}\label{sumsec}

\subsection{Overview: Interpreting the $X$ factor}\label{recapsec}

By performing radiative transfer calculations on model MCs, we assess
the physical properties which are most responsible for setting the
\Xfac, the ratio of molecular column density to integrated CO
intensity.  The fiducial model MC is a result of a magnetohydrodynamic
simulation of a gaseous medium responding to driven turbulence with a
treatment of chemistry which tracks the formation and destruction of
\Ht\ and CO; the resulting MC has many properties similar to those
found in Galactic MCs (Section \ref{modsec} and Paper I).  After
discussing the \Xfac\ from the fiducial model, we modify the physical
characteristics of the model MC and recompute the emerging CO line
emission from the altered models.  We then compare the modified
\Xfac\ with that provided by the original model (Section
\ref{ressec}), and discuss our results in the context of observed
trends (Section \ref{discsec}).

Our analysis was aimed at understanding the limited range in the
\X\ found from Galactic MC observations, shown in Figure
\ref{Xenvirons}.  We discussed the trend of a higher \Xfac\ for low
density regions in Paper I.  Basically, environments with lower CO
densities have lower CO intensities.  This could be due to lower total
density, higher background UV radiation fields, or lower
metallicities.  Though \siggas\ in the simulations ranges between
$\sim$50-500 \msunpc, and the \Xfac\ along individual LoSs can vary
significantly, in all except the very low metallicity models the
cloud-averaged $\langle X\rangle \sim$ \Xgal\ (see Fig. \ref{meanX}).
In this work, we have found that it is the limited range in velocities
and temperatures that constrains \W, and thereby \X, for a given
column density.

This interpretation applies for systems above a threshold column
density beyond which CO is optically thick ($\sim$20
\msun\ \pc$^{-2}$), but below very high column densities commonly
found in ULIRGs.  The precise value of the threshold column density
depends on a number of environmental factors of the source, such as
the metallicity, UV radiation intensity, and dust density.  Below this
column density, CO is optically thin, so a small decrease in column
density results in a stronger decrease in CO intensity \W, and thereby
an increase in \X.  At larger column densities, CO is optically thick.
Thus, an increase in CO abundance may or may not lead to an increase
in CO intensity, depending on the distribution of gas velocity.  Much
larger CO column densities are usually found in molecular rich sources
such as ULIRGs, which generally have higher temperatures, total
densities, as well as observed velocity dispersions (due to rotation
and/or turbulence) $-$ conditions which are not self-consistently
followed in the models presented here.  

An important consequence of our conclusion is that a given
\Xfac\ value may not always provide an accurate mass estimate of MCs
detected from CO observations.  Observationally, \Xfac\ relations
between total mass and CO brightness may coincidentally work well over
suitably large areas of molecular clouds.  This may simply reflect the
limited range in the mean column density, temperature, and velocity
dispersion within typical MCs.  These underlying global properties of
the emitting regions make a bridge between the average of many LoSs
with individual LoSs, where the relationships generally fail.  MC
masses estimated through a chosen constant \Xfac\ should only be
thought of as a first approximation of the gas mass, to within
$\sim$50\%.  Indeed, that derived \Xfac s for Galactic MCs themselves
vary by up to a factor of five has already been a warning sign that a
universal CO luminosity to gas mass conversion does not exist
\citep[e.g.][and
  Fig. \ref{Xenvirons}]{Solomonetal87,Okaetal98,Pinedaetal08}.  Due to
its high optical depth, there may be a significant amount of CO
self-absorption.  As a result, CO intensities only provide rough
estimates of the gas mass.

\subsection{Main Results}\label{sumlist}

1) We find a roughly constant \X\ throughout the model MC only when
the CO intensity is integrated over all velocities.  When intensities
at each PPV location are considered along with the column density
\NHtnu\ associated with that PPV position, there is a clear trend of
an increasing \Xfac\ with increasing column density.  This occurs
because in high density regions the CO line is saturated, so
increasing the column density only leads to more self-absorption,
resulting in an increase in the column density - CO intensity ratio
(Section 4.1).

2) The variation in \X\ between different LoSs through an individual
cloud is not due to the variation in temperature within the MC.  When
setting the temperature to the mass-weighted average value, the
original \Xfac\ distribution is recovered.  The cloud-averaged
\Xfac\ is thus determined primarily by the densest regions in the MC.
We find a weak dependence of \X\ on $T$: roughly $\langle X \rangle
\propto T^{-0.5}$ from 20 K to 100 K if density and velocity structure
are fixed (Section 4.2).

3) Even though the \Xfac\ is defined with a column density, it also
depends on the volume density \nHt.  This is due in part to the
saturation of the CO line at high densities, as indicated in 1) above.
Increasing the mean density from $n_0=300$ to $1000$ \cmt only
decreases $\langle X\rangle$ by 15\%, however (Section 4.3).

4) Similarly, at high CO abundances \X\ increases with \NHt\ due to
line saturation.  At very low abundances (\fCO \aplt $10^{-6}$), the
line is optically thin, so \W\ is directly proportional to
\NHt\ producing a constant \Xfac.  Since there is a range in CO
abundances in MCs (\citealt{Gloveretal10}, Paper I), the \Xfac\ shows
signatures of both high (saturation) and low opacities (Section 4.3).
On average, the \Xfac\ is not strongly dependent on the \fCO\ if
\fCO\ is in the observed range $\in 10^{-5} - 10^{-4}$ (Section 4.3).

5) The fiducial model MC has a linewidth-size relationship $\sigma
\propto R^{1/2}$, and produces \X $\sim$\Xgal, similar to
observations.  Nevertheless, the \Xfac\ from models without any
linewidth-size relationship, but with a similar range in velocities,
reproduces the \Xfac\ trends in the fiducial model.  Further, in
models with a larger range in velocities, there is less radiation
trapping, leading to larger CO intensities and correspondingly lower
\Xfac s.  Thus, it is the {\it range} in velocities that determine the
\Xfac.  The details of the velocity structure do not influence
\X\ (Section 4.4).  Since the CO brightness temperature also depends
on $\sigma$, the \Xfac\ does not decrease linearly with increasing
$\sigma$, but instead shows a variation closer to $X \propto
\sigma^{-1/2}$ for $\sigma=2-20$ \kms (Section 4.6).

6) A CO-bright LoS with a large linewidth may be due to a single high
density region.  Alternatively, a similar line profile may result from
a LoS with numerous lower density gaseous peaks spanning a range of
velocities.  Therefore, two LoSs may have similar integrated
intensities, but vastly different morphologies.  Due to this diversity
in morphology in turbulent clouds and the optically thick nature of
CO, CO emission does not trace \Ht\ in a one-to-one fashion on
individual LoSs.  The ``mist'' model
\citep{Dickmanetal86,Solomonetal87} does not accurately capture the
complexity of turbulence and radiative transfer of the CO line in MCs
(Section 4.5).

7) Similar to individual LoSs, cloud-averaged spectra do not
represent the total molecular content in a one-to-one fashion.  Due to
the blending of numerous emitting features, the averaged CO line is
Gaussian.  However, in experiments where some gas is removed, the
intensity of the resulting spectra is not simply scaled down by the
factor that would be appropriate if CO were a linear tracer of
molecular gas (Section 4.6).

8) Since we find that the \Xfac\ is insensitive to the detailed
velocity structure, we argue that the reason Galactic MC observations
generally find \X $\sim$ \Xgal\ is due to the limited {\it range} in
observed linewidths $\sim$ 2-5 \kms, as well as in the temperatures
within dense, CO-abundant regions in MCs ($\sim$10-25 K, Section 5.1),
and in the column densities.  Thus, the nearly constant \Xfac\ found
in Galactic MCs does not necessarily require that $\sigma \propto
R^{1/2}$, or that clouds are ``virialized.''  Although a near-constant
\MtoL\ ratio would result if the velocities were governed solely by
gas self-gravity, and $\sigma \propto R^{1/2}$ universally, these are
not necessary conditions for \X\ to be nearly constant, and may or may
not be true for Galactic MCs (Section 5.2).

9) Given our explanation that the approximately constant
\X$\sim$\Xgal\ found in Galactic MCs is due to the limited range in MC
properties, masses deduced through a constant \X\ should only be
considered as a rough estimate, within a factor $\sim 2$.  Taking our
results from Paper I and the current work together, we find
cloud-averaged $X\sim 2-4 \times10^{20}$\Xunits\ for Solar-metallicity
models with $T\sim 10-20$ K, velocity dispersions $\sigma \sim 1-6$
\kms, and $=N_{H_2} \sim 2-20 \times 10^{21}$\cmsq -- i.e. similar to
Galactic MCs.  The larger \Xfac\ values found in lower metallicity
systems can be understood due to lower CO abundances (see references
in Section 5.4).  In molecular-dominated regions such as ULIRGs, the
combined effect of higher CO and \Ht\ volume and column densities,
higher temperatures (due to enhanced star formation), and larger
velocities (due to rotation and/or turbulence) all need to be
self-consistently treated to assess the \Xfac.

\section*{Acknowledgements}
We are grateful to Desika Narayanan, Alyssa Goodman, Jaime Pineda,
Mordecai Mac Low, and Jerome Pety for useful discussions on CO
emission.  We also thank the anonymous referee for suggestions that
improved our presentation.  Several of the simulations reported on in
this paper were performed on the Ranger cluster at the Texas Advanced
Computing Center using time allocated as part of Teragrid project
TG-MCA99S024.  R.S.K. acknowledges financial support from the
Landesstiftung Baden-W\"urrtemberg via their program International
Collaboration II (grant P-LS-SPII/18), and subsidies from the DFG
under grants no. KL1358/1, KL1358/4, KL1358/5, KL1358/10, and
KL1358/11, as well as from a Frontier grant of Heidelberg University
sponsored by the German Excellence Initiative. R.S., S.C.G., and
R.S.K. are supported by the German Bundesministerium f\"ur Bildung und
Forschung via the ASTRONET project STAR FORMAT (grant 05A09VHA).  This
work was supported in part by the U.S. Department of Energy contract
no. DE-AC-02-76SF00515.  E.C.O. acknowledges support from grant AST
0908185 from the U.S. National Science Foundation.

\appendix
\section{The Effect of Microturbulence}

As indicated in Section \ref{modsec} for the radiative transfer
calculation, we set the microturbulent velocity \vmt = 0.5 \kms\ in
each zone of the simulation.  This value is motivated by loosely
extrapolating the observed linewidth-size relationship
\citep[][]{Larson81,Solomonetal87} down to the physical resolution of
the simulation ($\sim$0.1 pc).  Due to observational uncertainties and
the scatter in the observed linewidth-size relationship, the turbulent
velocity at any given scale is not precisely known.  We have thus
explored how other values of \vmt\ may affect our results.

\begin{figure}
\includegraphics[width=90mm]{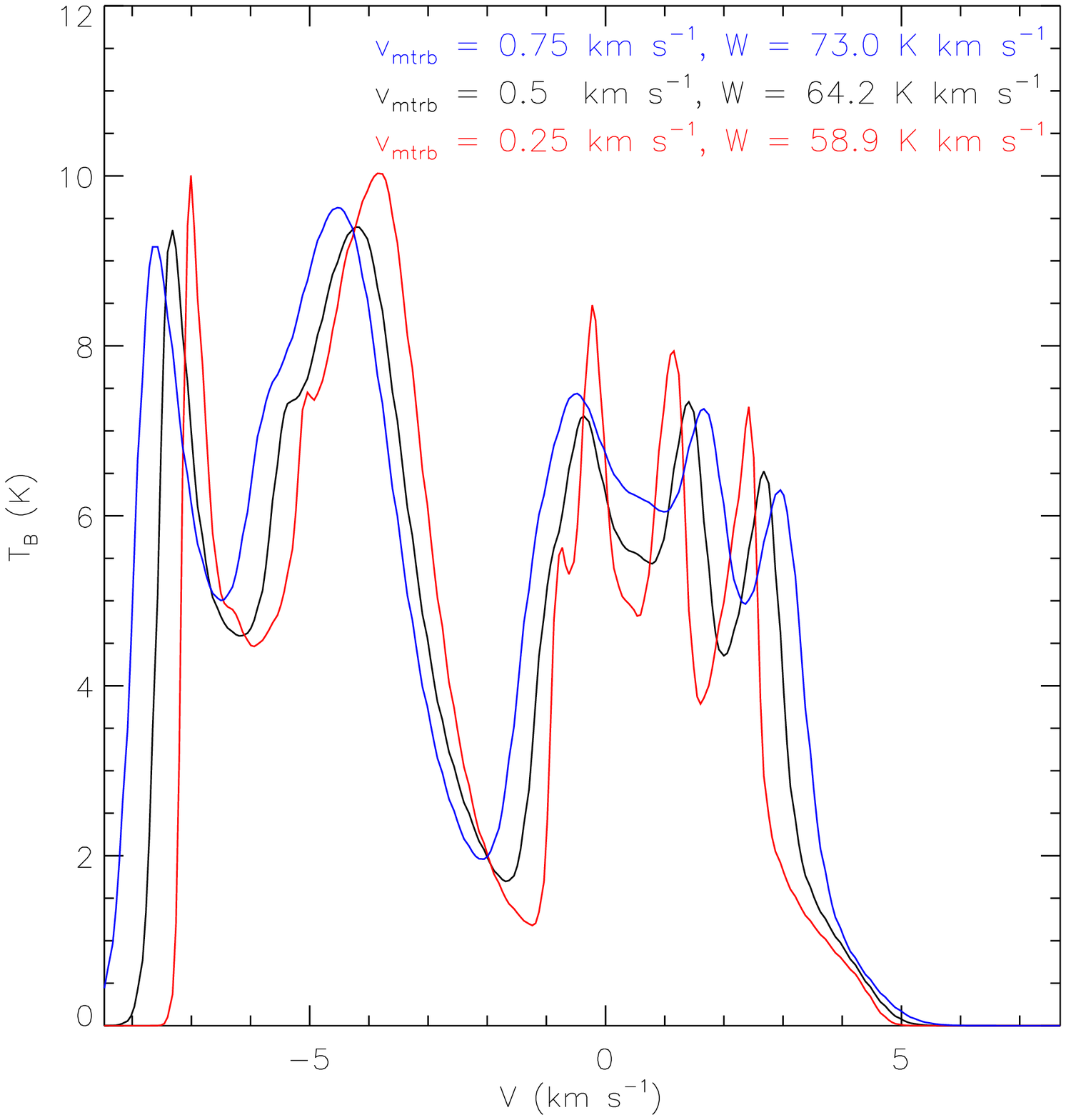}
\caption{CO spectra from LoS B (see Fig. \ref{spec1}).  Three
  different microturbulent velocities are considered: 0.25 (red), 0.5
  (black) and 0.75 (blue) \kms.  The velocity-integrated intensities
  \W\ are indicated.}
\label{turbeff}
\end{figure}

Figure \ref{turbeff} shows the spectrum from one LoS.  It corresponds
to LoS B shown in Figure \ref{spec1}.  Three spectra are shown, with
different values of \vmt = 0.25, 0.5, and 0.75 \kms.  With higher
microturbulence the line is broadened, and gaps or sharp features in
the spectrum can become smooth, as evident near velocities of $-$0.5
and $-$5.0 \kms.

The variations in the spectra lead to slight differences in velocity
integrated intensity \W.  For \vmt = 0.25, 0.5, and 0.75 \kms, \W\ =
58.9, 64.2, and 73.0 \Kkms.  The differences in integrated intensity
will of course result in corresponding differences in the \Xfac.  For
this LoS, the column density is 1.1$\times 10^{22}$ \cmsq, providing
\X\ = 1.9, 1.7, and 1.5 $\times 10^{20}$ \Xunits.  In the
cloud-averaged spectra (e.g. Fig. \ref{gausfits}), the smooth Gaussian
profiles are not significantly affected, and the linewidths only vary
by \aplt 6\% from the fiducial model (e.g. Fig. \ref{gausfits}).  For
\vmt = 0.25, 0.5, and 0.75, the peak $T_{B,0}$= 5.1, 5.8, and 6.3 K,
leading to cloud-averaged \Xfac s 2.6, 2.2, 1.9$\times 10^{20}$
\Xunits, respectively.

We have found that the precise value of \vmt\ does not affect the
cloud-averaged spectrum $-$ and the \Xfac\ $-$ significantly.  The
most noticeable difference occurs in the line shapes along individual
LoSs.  Nevertheless, even then the differences are minimal.  Any value
of \vmt\ within the reasonable range $\in 0.25 - 0.75$ would produce
very similar CO intensities.  Thus, the values of the \Xfac\ found in
this work, and the conclusions drawn from its analysis, would not be
affected if different values of \vmt\ were employed.

\section{Numerical Resolution}

While performing radiative transfer calculations, there is a potential
for obtaining inaccurate line intensities due to unresolved velocity
structure.  In this section, we discuss why resolving the velocity
gradients is important, and how we have accounted for under-resolved
velocities in our study.

In a grid-based model, the physical properties are only defined in
each zone - either at the zone center or zone edge.  While integrating
the equation of radiative transfer, or ray-tracing, through a given
LoS, the optical depth is only updated when stepping from one grid
zone to the next.  For a given ``observed'' frequency $\nu$,
corresponding to velocity $v_{\rm obs}$, the optical depth $\tau_\nu$
is only modified in zones with LoS velocities that correspond to the
non-zero portion of the Gaussian profile function $\phi_\nu$ (see
Eqn. \ref{taueqn} and Eqn. 6 in Paper I).

\begin{figure}
\includegraphics[width=90mm]{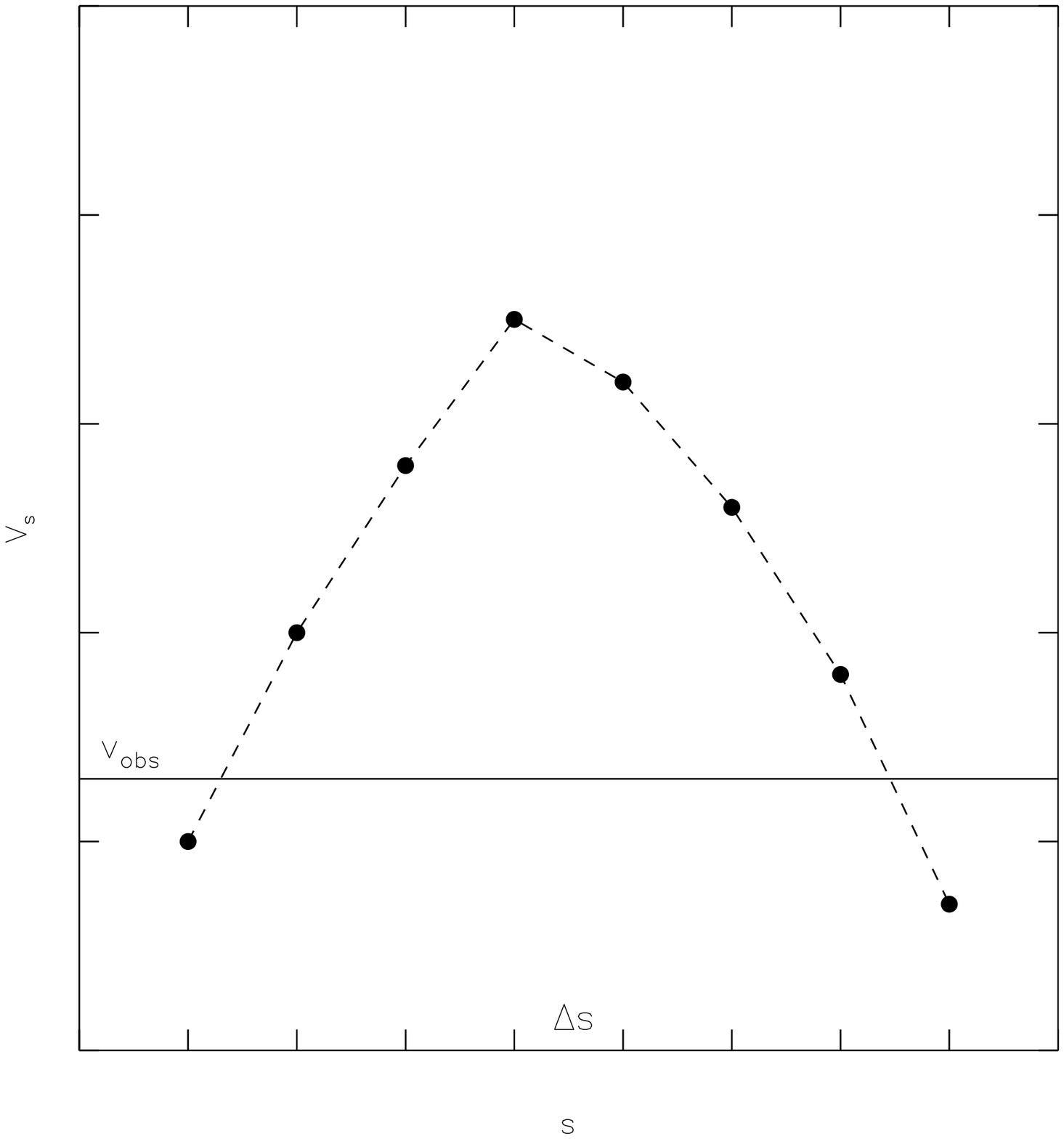}
\caption{Velocity component along the LoS, $v_s$ vs. $s$.  A
  sample observed frequency (or velocity) is shown by the solid line,
  $v_{obs}$.  The units of $v_s$ and $s$ are arbitrary.  Due to the
  large velocity gradients, there would not be any observed emission
  at this $v_{\rm obs}$, unless the velocities were interpolated
  (dashed line).}
\label{vs}
\end{figure}

As an example, consider the velocity profile of a sample LoS shown in
Figure \ref{vs}.  The LoS component of the velocity $v_s$ is only
defined at discrete locations along the LoS $s$ (e.g. edges of grid
zones).  At the indicated $v_{\rm obs}$, due to the large velocity
gradients, there would be no emission (or, if optically thick,
absorption) from this region, since none of the velocities overlap
with $v_{\rm obs}$.  Of course, if the (Gaussian) width of $\phi_\nu$
were large enough (e.g. due to microturbulence), then there may be a
contribution from one or more zones.  For our present discussion,
however, we will assume that $\phi_\nu$ has a narrow width (e.g. the
vertical extent of the dots defining $v_s$ in Fig. \ref{vs}).

In MCs, and generally in any environment, the velocities are not
discretized but rather vary continuously throughout the medium.
Therefore, the velocities in Figure \ref{vs} are more accurately
described by the dashed line connecting the defined $v_s$s in each
zone.  Consequently, at the observed velocity there would in fact be
emission along this LoS, due to the overlap of $v_s$ and $v_{\rm obs}$
(at two locations in this example).

In order to account for the continuous variation in cloud properties,
we have implemented interpolation into our radiative transfer
calculations \citep[also see][]{Pontoppidanetal09}.  All the required
quantities, such as temperature, CO density, \Ht\ density, and
position, are linearly interpolated, effectively increasing the
physical resolution along the LoS.

\begin{figure*}
\includegraphics[width=120mm]{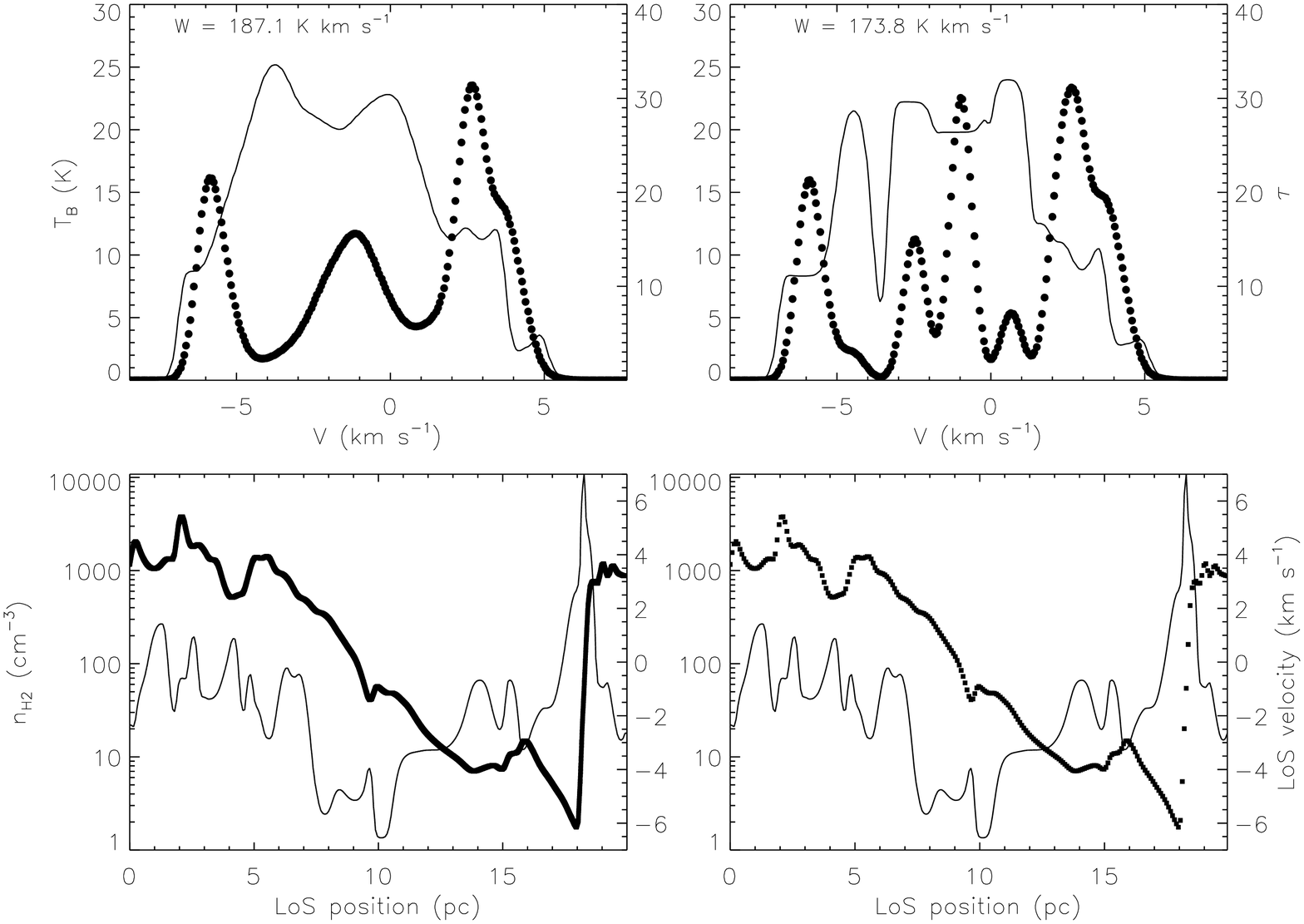}
\caption{Spectrum, densities and velocities from LoS A (as in
  Fig. \ref{spec1}).  Top panel shows the brightness temperature
  (lines - left axis), and integrated optical depth (circles - right
  axis) as a function of observed velocity.  Bottom panels show
  density (lines - left axis) and LoS velocity (squares - right axis)
  as a function of LoS position.  Left panels show quantities derived
  using interpolation, and right panels show the quantities derived
  directly from the original 256$^3$ simulation.  The integrated
  intensities $W$ are indicated in the top panels.  }
\label{ss}
\end{figure*}

Figure \ref{ss} shows how under-sampling the velocities may produce
inaccurate line profiles.  The figure shows the relevant quantities
from LoS A shown in Figure \ref{spec1} and discussed in Section
\ref{specsec}.  The left panels show the results using interpolation
(16$\times$ the original grid, so $\Delta s$ = 0.0049 pc), and the
right panels show the result using the original simulation grid
($\Delta s$ = 0.078 pc).  Clearly, there are stark differences in the
spectra.  When unresolved, the spectrum has numerous ``flat tops,''
and there is a sharp decrease in the intensities between $-3$ and
$-4$ \kms.  These features do not occur in the spectrum obtained from
the interpolated LoS.  The ``flat tops'' and intensity drop could be
misidentified as saturation or absorption features, respectively, when
in fact they are simply features arising due to the under-resolved
grid.

The differences can be understood from the velocity and optical depth
profiles.  In the high density region responsible for most of the
emission (see Section \ref{specsec}), there is a very large velocity
gradient.  At an observed velocity of $-4$ \kms, there would not be
any emission from this region, due to the large gap in the velocities
at LoS position $\sim$18.5 pc.  However, when interpolated (left
panels), there are continuous velocities in this region, resulting in
enhanced emission at $-4$ \kms.

For the interpolated 0.0049 pc and original 0.078 pc simulations, the
integrated intensities from LoS A are 187.1 and 173.8 \Kkms,
respectively.  This results in a slight difference of $\sim$10 \%, and
would lead to a similar discrepancy in the \Xfac\ from this LoS.  The
precise difference would of course depend on the particular LoS $-$
LoSs with many regions with large velocity gradients would be strongly
affected by under-resolved velocities.  For model n300, the
cloud-average \Xfac\ is $\langle X \rangle $=2.2$\times 10^{20}$
\Xunits, regardless of whether interpolation is employed.  Thus, for
the n300 model, only the line shapes along individual LoS are affected
by under-resolution, but the velocities are sufficiently resolved for
estimating the cloud-averaged integrated intensity.

For the models discussed in Section 4.4 where we have directly
modified the velocities to have larger dispersions, there will of
course be large velocity gradients in adjacent zones.  For these
models, we find that interpolation results in more significant
differences.  Recall that in the altered models, the velocities are
replaced in the original grid utilized in the chemo-MHD simulation.
The altered velocities on this grid have the chosen dispersions.  In
the radiative transfer calculations, due to the large velocity
gradients, interpolation leads to the addition of numerous values to
the velocity dataset.  In the models with large chosen dispersions in
the original grid, this interpolation can potentially lead to a
velocity dispersion that is different from the chosen one.  For
instance, in the $\sigma = $ 24 and 240 \kms\ models, after
interpolation (with 16$\times$ increased resolution), the effective
velocity dispersion becomes 21 and 205 \kms, respectively.  In both
the $\sigma =$ 24 and 240 \kms\ models, we find that interpolation
increases the cloud-averaged $W$ by factors of 2 and 4, respectively,
and thereby decreases the \Xfac\ by a corresponding amount.  As these
models are highly artificial to begin with, we do not ensure that all
the velocities are well resolved.  Nevertheless, qualitatively, we can
be sure that increasing the velocity dispersion leads to decreasing
the \Xfac.  As we state in Section 5.3, models of real sources with
very high dispersions, such as ULIRGs, would have to be well resolved
in order to quantify the \Xfac$-\sigma$ relationship.

\bibliography{citations}

\begin{thebibliography}{}

\bibitem[\protect\citeauthoryear{{Ballesteros-Paredes}}{{Ballesteros-Paredes}}%
{2006}]{Ballesteros-Paredes06}
{Ballesteros-Paredes} J.,  2006, \mnras, 372, 443

\bibitem[\protect\citeauthoryear{{Ballesteros-Paredes} \& {Mac
  Low}}{{Ballesteros-Paredes} \& {Mac
  Low}}{2002}]{Ballesteros-Paredes&MacLow02}
{Ballesteros-Paredes} J.,  {Mac Low} M.-M.,  2002, \apj, 570, 734

\bibitem[\protect\citeauthoryear{{Bertoldi} \& {McKee}}{{Bertoldi} \&
  {McKee}}{1992}]{Bertoldi&McKee92}
{Bertoldi} F.,  {McKee} C.~F.,  1992, \apj, 395, 140

\bibitem[\protect\citeauthoryear{{Bolatto}, {Leroy}, {Rosolowsky}, {Walter} \&
  {Blitz}}{{Bolatto} et~al.}{2008}]{Bolattoetal08}
{Bolatto} A.~D.,  {Leroy} A.~K.,  {Rosolowsky} E.,  {Walter} F.,    {Blitz} L.,
   2008, \apj, 686, 948

\bibitem[\protect\citeauthoryear{{Boselli}, {Gavazzi}, {Lequeux}, {Buat},
  {Casoli}, {Dickey} \& {Donas}}{{Boselli} et~al.}{1997}]{Bosellietal97}
{Boselli} A.,  {Gavazzi} G.,  {Lequeux} J.,  {Buat} V.,  {Casoli} F.,  {Dickey}
  J.,    {Donas} J.,  1997, \aap, 327, 522

\bibitem[\protect\citeauthoryear{{Boselli}, {Lequeux} \& {Gavazzi}}{{Boselli}
  et~al.}{2002}]{Bosellietal02}
{Boselli} A.,  {Lequeux} J.,    {Gavazzi} G.,  2002, \aap, 384, 33

\bibitem[\protect\citeauthoryear{{Burgh}, {France} \& {McCandliss}}{{Burgh}
  et~al.}{2007}]{Burghetal07}
{Burgh} E.~B.,  {France} K.,    {McCandliss} S.~R.,  2007, \apj, 658, 446

\bibitem[\protect\citeauthoryear{{Combes}}{{Combes}}{1991}]{Combes91}
{Combes} F.,  1991, \araa, 29, 195

\bibitem[\protect\citeauthoryear{{Dame}, {Hartmann} \& {Thaddeus}}{{Dame}
  et~al.}{2001}]{Dameetal01}
{Dame} T.~M.,  {Hartmann} D.,    {Thaddeus} P.,  2001, \apj, 547, 792

\bibitem[\protect\citeauthoryear{{Dickman}}{{Dickman}}{1978}]{Dickman78}
{Dickman} R.~L.,  1978, \apjs, 37, 407

\bibitem[\protect\citeauthoryear{{Dickman}, {Snell} \& {Schloerb}}{{Dickman}
  et~al.}{1986}]{Dickmanetal86}
{Dickman} R.~L.,  {Snell} R.~L.,    {Schloerb} F.~P.,  1986, \apj, 309, 326

\bibitem[\protect\citeauthoryear{{Downes} \& {Solomon}}{{Downes} \&
  {Solomon}}{1998}]{Downes&Solomon98}
{Downes} D.,  {Solomon} P.~M.,  1998, \apj, 507, 615

\bibitem[\protect\citeauthoryear{{Downes}, {Solomon} \& {Radford}}{{Downes}
  et~al.}{1993}]{Downesetal93}
{Downes} D.,  {Solomon} P.~M.,    {Radford} S.~J.~E.,  1993, \apjl, 414, L13

\bibitem[\protect\citeauthoryear{{Draine}}{{Draine}}{1978}]{Draine78}
{Draine} B.~T.,  1978, \apjs, 36, 595

\bibitem[\protect\citeauthoryear{{Gammie}, {Lin}, {Stone} \&
  {Ostriker}}{{Gammie} et~al.}{2003}]{Gammieetal03}
{Gammie} C.~F.,  {Lin} Y.-T.,  {Stone} J.~M.,    {Ostriker} E.~C.,  2003, \apj,
  592, 203

\bibitem[\protect\citeauthoryear{{Glover}, {Federrath}, {Mac Low} \&
  {Klessen}}{{Glover} et~al.}{2010}]{Gloveretal10}
{Glover} S.~C.~O.,  {Federrath} C.,  {Mac Low} M.,    {Klessen} R.~S.,  2010,
  \mnras, 404, 2

\bibitem[\protect\citeauthoryear{{Glover} \& {Mac Low}}{{Glover} \& {Mac
  Low}}{2007a}]{Glover&MacLow07I}
{Glover} S.~C.~O.,  {Mac Low} M.,  2007a, \apjs, 169, 239

\bibitem[\protect\citeauthoryear{{Glover} \& {Mac Low}}{{Glover} \& {Mac
  Low}}{2007b}]{Glover&MacLow07II}
{Glover} S.~C.~O.,  {Mac Low} M.,  2007b, \apj, 659, 1317

\bibitem[\protect\citeauthoryear{{Glover} \& {Mac Low}}{{Glover} \& {Mac
  Low}}{2011}]{Glover&MacLow11}
{Glover} S.~C.~O.,  {Mac Low} M.,  2011, \mnras, 412, 337

\bibitem[\protect\citeauthoryear{{Grenier}, {Casandjian} \&
  {Terrier}}{{Grenier} et~al.}{2005}]{Grenieretal05}
{Grenier} I.~A.,  {Casandjian} J.,    {Terrier} R.,  2005, Science, 307, 1292

\bibitem[\protect\citeauthoryear{{Habing}}{{Habing}}{1968}]{Habing68}
{Habing} H.~J.,  1968, \bain, 19, 421

\bibitem[\protect\citeauthoryear{{Heyer}, {Krawczyk}, {Duval} \&
  {Jackson}}{{Heyer} et~al.}{2009}]{Heyeretal09}
{Heyer} M.,  {Krawczyk} C.,  {Duval} J.,    {Jackson} J.~M.,  2009, \apj, 699,
  1092

\bibitem[\protect\citeauthoryear{{Israel}}{{Israel}}{1997}]{Israel97}
{Israel} F.~P.,  1997, \aap, 328, 471

\bibitem[\protect\citeauthoryear{{Israel}, {de Graauw}, {van de Stadt} \& {de
  Vries}}{{Israel} et~al.}{1986}]{Israeletal86}
{Israel} F.~P.,  {de Graauw} T.,  {van de Stadt} H.,    {de Vries} C.~P.,
  1986, \apj, 303, 186

\bibitem[\protect\citeauthoryear{{Kauffmann}, {Pillai}, {Shetty}, {Myers} \&
  {Goodman}}{{Kauffmann} et~al.}{2010a}]{Kauffmannetal10a}
{Kauffmann} J.,  {Pillai} T.,  {Shetty} R.,  {Myers} P.~C.,    {Goodman} A.~A.,
   2010a, \apj, 712, 1137

\bibitem[\protect\citeauthoryear{{Kauffmann}, {Pillai}, {Shetty}, {Myers} \&
  {Goodman}}{{Kauffmann} et~al.}{2010b}]{Kauffmannetal10b}
{Kauffmann} J.,  {Pillai} T.,  {Shetty} R.,  {Myers} P.~C.,    {Goodman} A.~A.,
   2010b, \apj, 716, 433

\bibitem[\protect\citeauthoryear{{Kutner} \& {Leung}}{{Kutner} \&
  {Leung}}{1985}]{Kutner&Leung85}
{Kutner} M.~L.,  {Leung} C.~M.,  1985, \apj, 291, 188

\bibitem[\protect\citeauthoryear{{Larson}}{{Larson}}{1981}]{Larson81}
{Larson} R.~B.,  1981, \mnras, 194, 809

\bibitem[\protect\citeauthoryear{{Leroy}, {Bolatto}, {Stanimirovic}, {Mizuno},
  {Israel} \& {Bot}}{{Leroy} et~al.}{2007}]{Leroyetal07}
{Leroy} A.,  {Bolatto} A.,  {Stanimirovic} S.,  {Mizuno} N.,  {Israel} F.,
  {Bot} C.,  2007, \apj, 658, 1027

\bibitem[\protect\citeauthoryear{{Leroy}, {Bolatto}, {Bot}, {Engelbracht},
  {Gordon}, {Israel}, {Rubio}, {Sandstrom} \& {Stanimirovi{\'c}}}{{Leroy}
  et~al.}{2009}]{Leroyetal09}
{Leroy} A.~K.,  {Bolatto} A.,  {Bot} C.,  {Engelbracht} C.~W.,  {Gordon} K.,
  {Israel} F.~P.,  {Rubio} M.,  {Sandstrom} K.,    {Stanimirovi{\'c}} S.,
  2009, \apj, 702, 352

\bibitem[\protect\citeauthoryear{{Leroy}, {Bolatto}, {Gordon}, {Sandstrom},
  {Gratier}, {Rosolowsky}, {Engelbracht}, {Mizuno}, {Corbelli}, {Fukui} \&
  {Kawamura}}{{Leroy} et~al.}{2011}]{Leroyetal11}
{Leroy} A.~K.,  {Bolatto} A.,  {Gordon} K.,  {Sandstrom} K.,  {Gratier} P.,
  {Rosolowsky} E.,  {Engelbracht} C.~W.,  {Mizuno} N.,  {Corbelli} E.,  {Fukui}
  Y.,    {Kawamura} A.,  2011, ArXiv e-prints

\bibitem[\protect\citeauthoryear{{Liszt}, {Pety} \& {Lucas}}{{Liszt}
  et~al.}{2010}]{Lisztetal10}
{Liszt} H.~S.,  {Pety} J.,    {Lucas} R.,  2010, \aap, 518, A45+

\bibitem[\protect\citeauthoryear{{Lombardi}, {Alves} \& {Lada}}{{Lombardi}
  et~al.}{2006}]{Lombardietal06}
{Lombardi} M.,  {Alves} J.,    {Lada} C.~J.,  2006, \aap, 454, 781

\bibitem[\protect\citeauthoryear{{Mac Low} \& {Klessen}}{{Mac Low} \&
  {Klessen}}{2004}]{MacLow&Klessen04}
{Mac Low} M.,  {Klessen} R.~S.,  2004, Reviews of Modern Physics, 76, 125

\bibitem[\protect\citeauthoryear{{Maloney}}{{Maloney}}{1990}]{Maloney90}
{Maloney} P.,  1990, \apjl, 348, L9

\bibitem[\protect\citeauthoryear{{Maloney} \& {Black}}{{Maloney} \&
  {Black}}{1988}]{Maloney&Black88}
{Maloney} P.,  {Black} J.~H.,  1988, \apj, 325, 389

\bibitem[\protect\citeauthoryear{{McKee} \& {Ostriker}}{{McKee} \&
  {Ostriker}}{2007}]{McKee&Ostriker07}
{McKee} C.~F.,  {Ostriker} E.~C.,  2007, \araa, 45, 565

\bibitem[\protect\citeauthoryear{{Myers} \& {Goodman}}{{Myers} \&
  {Goodman}}{1988}]{Myers&Goodman88}
{Myers} P.~C.,  {Goodman} A.~A.,  1988, \apj, 329, 392

\bibitem[\protect\citeauthoryear{{Oka}, {Hasegawa}, {Hayashi}, {Handa} \&
  {Sakamoto}}{{Oka} et~al.}{1998}]{Okaetal98}
{Oka} T.,  {Hasegawa} T.,  {Hayashi} M.,  {Handa} T.,    {Sakamoto} S.,  1998,
  \apj, 493, 730

\bibitem[\protect\citeauthoryear{{Pineda}, {Caselli} \& {Goodman}}{{Pineda}
  et~al.}{2008}]{Pinedaetal08}
{Pineda} J.~E.,  {Caselli} P.,    {Goodman} A.~A.,  2008, \apj, 679, 481

\bibitem[\protect\citeauthoryear{{Planck Collaboration}, {Ade}, {Aghanim},
  {Arnaud}, {Ashdown}, {Aumont}, {Baccigalupi}, {Balbi}, {Banday}, {Barreiro}
  \& et al.}{{Planck Collaboration} et~al.}{2011}]{Planck11}
{Planck Collaboration} {Ade} P.~A.~R.,  {Aghanim} N.,  {Arnaud} M.,  {Ashdown}
  M.,  {Aumont} J.,  {Baccigalupi} C.,  {Balbi} A.,  {Banday} A.~J.,
  {Barreiro} R.~B.,    et al. 2011, ArXiv e-prints 1101.2029

\bibitem[\protect\citeauthoryear{{Polk}, {Knapp}, {Stark} \& {Wilson}}{{Polk}
  et~al.}{1988}]{Polketal88}
{Polk} K.~S.,  {Knapp} G.~R.,  {Stark} A.~A.,    {Wilson} R.~W.,  1988, \apj,
  332, 432

\bibitem[\protect\citeauthoryear{{Pontoppidan}, {Meijerink}, {Dullemond} \&
  {Blake}}{{Pontoppidan} et~al.}{2009}]{Pontoppidanetal09}
{Pontoppidan} K.~M.,  {Meijerink} R.,  {Dullemond} C.~P.,    {Blake} G.~A.,
  2009, \apj, 704, 1482

\bibitem[\protect\citeauthoryear{{Roman-Duval}, {Jackson}, {Heyer}, {Rathborne}
  \& {Simon}}{{Roman-Duval} et~al.}{2010}]{Roman-Duvaletal10}
{Roman-Duval} J.,  {Jackson} J.~M.,  {Heyer} M.,  {Rathborne} J.,    {Simon}
  R.,  2010, \apj, 723, 492

\bibitem[\protect\citeauthoryear{{Rosolowsky}, {Pineda}, {Kauffmann} \&
  {Goodman}}{{Rosolowsky} et~al.}{2008}]{Rosolowskyetal08}
{Rosolowsky} E.~W.,  {Pineda} J.~E.,  {Kauffmann} J.,    {Goodman} A.~A.,
  2008, \apj, 679, 1338

\bibitem[\protect\citeauthoryear{{Sch{\"o}ier}, {van der Tak}, {van Dishoeck}
  \& {Black}}{{Sch{\"o}ier} et~al.}{2005}]{Schoieretal05}
{Sch{\"o}ier} F.~L.,  {van der Tak} F.~F.~S.,  {van Dishoeck} E.~F.,    {Black}
  J.~H.,  2005, \aap, 432, 369

\bibitem[\protect\citeauthoryear{{Shetty}, {Collins}, {Kauffmann}, {Goodman},
  {Rosolowsky} \& {Norman}}{{Shetty} et~al.}{2010}]{Shettyetal10}
{Shetty} R.,  {Collins} D.~C.,  {Kauffmann} J.,  {Goodman} A.~A.,  {Rosolowsky}
  E.~W.,    {Norman} M.~L.,  2010, \apj, 712, 1049

\bibitem[\protect\citeauthoryear{{Shetty}, {Glover}, {Dullemond} \&
  {Klessen}}{{Shetty} et~al.}{2011}]{Shettyetal11}
{Shetty} R.,  {Glover} S.~C.,  {Dullemond} C.~P.,    {Klessen} R.~S.,  2011,
  \mnras, 412, 1686

\bibitem[\protect\citeauthoryear{{Sobolev}}{{Sobolev}}{1957}]{Sobolev57}
{Sobolev} V.~V.,  1957, Soviet Astronomy, 1, 678

\bibitem[\protect\citeauthoryear{{Solomon}, {Downes}, {Radford} \&
  {Barrett}}{{Solomon} et~al.}{1997}]{Solomonetal97}
{Solomon} P.~M.,  {Downes} D.,  {Radford} S.~J.~E.,    {Barrett} J.~W.,  1997,
  \apj, 478, 144

\bibitem[\protect\citeauthoryear{{Solomon}, {Rivolo}, {Barrett} \&
  {Yahil}}{{Solomon} et~al.}{1987}]{Solomonetal87}
{Solomon} P.~M.,  {Rivolo} A.~R.,  {Barrett} J.,    {Yahil} A.,  1987, \apj,
  319, 730

\bibitem[\protect\citeauthoryear{{Solomon} \& {Vanden Bout}}{{Solomon} \&
  {Vanden Bout}}{2005}]{SolomonVandenBout05}
{Solomon} P.~M.,  {Vanden Bout} P.~A.,  2005, \araa, 43, 677

\bibitem[\protect\citeauthoryear{{Strong}, {Bloemen}, {Dame}, {Grenier},
  {Hermsen}, {Lebrun}, {Nyman}, {Pollock} \& {Thaddeus}}{{Strong}
  et~al.}{1988}]{Strongetal88}
{Strong} A.~W.,  {Bloemen} J.~B.~G.~M.,  {Dame} T.~M.,  {Grenier} I.~A.,
  {Hermsen} W.,  {Lebrun} F.,  {Nyman} L.,  {Pollock} A.~M.~T.,    {Thaddeus}
  P.,  1988, \aap, 207, 1

\bibitem[\protect\citeauthoryear{{Tacconi}, {Genzel}, {Smail}, {Neri},
  {Chapman}, {Ivison}, {Blain}, {Cox} \& {Omont} A.}{{Tacconi}
  et~al.}{2008}]{Tacconietal08}
{Tacconi} L.~J.,  {Genzel} R.,  {Smail} I.,  {Neri} R.,  {Chapman} S.~C.,
  {Ivison} R.~J.,  {Blain} A.,  {Cox} P.,    {Omont} A. e.~a.,  2008, \apj,
  680, 246

\bibitem[\protect\citeauthoryear{{van Dishoeck} \& {Black}}{{van Dishoeck} \&
  {Black}}{1988}]{vanDishoeck&Black88}
{van Dishoeck} E.~F.,  {Black} J.~H.,  1988, \apj, 334, 771

\bibitem[\protect\citeauthoryear{{Wolfire}, {Hollenbach} \& {McKee}}{{Wolfire}
  et~al.}{2010}]{Wolfireetal10}
{Wolfire} M.~G.,  {Hollenbach} D.,    {McKee} C.~F.,  2010, \apj, 716, 1191

\bibitem[\protect\citeauthoryear{{Wolfire}, {Hollenbach} \&
  {Tielens}}{{Wolfire} et~al.}{1993}]{Wolfireetal93}
{Wolfire} M.~G.,  {Hollenbach} D.,    {Tielens} A.~G.~G.~M.,  1993, \apj, 402,
  195

\bibitem[\protect\citeauthoryear{{Yang}, {Stancil}, {Balakrishnan} \&
  {Forrey}}{{Yang} et~al.}{2010}]{Yangetal10}
{Yang} B.,  {Stancil} P.~C.,  {Balakrishnan} N.,    {Forrey} R.~C.,  2010,
  \apj, 718, 1062

\bibitem[\protect\citeauthoryear{{Young} \& {Scoville}}{{Young} \&
  {Scoville}}{1991}]{Young&Scoville91}
{Young} J.~S.,  {Scoville} N.~Z.,  1991, \araa, 29, 581

\end{thebibliography}

\label{lastpage}
\end{document}